\documentclass[oupdraft]{bio}
\usepackage{xcolor}
\usepackage{float}

\DeclareMathOperator*{\argmax}{arg\,max}
\DeclareMathOperator{\E}{\text{E}}
\DeclareMathOperator{\Var}{\text{Var}}
\newtheorem{proposition}{Proposition}
\newtheorem{algorithm}{Algorithm}

\usepackage{xr}
\begin{document}
\vspace{-10mm}
\title{Inference after latent variable estimation for single-cell RNA sequencing data}

\author{Anna Neufeld $^{\dagger}$, Lucy L. Gao $\circ$, Joshua Popp$^\bigtriangleup$, Alexis Battle$^{\bigtriangleup \bigtriangledown, \cdot}$, and Daniela Witten $^{\dagger\ddagger}$ \\[4pt]
$\dagger$ \textit{Department of Statistics, University of Washington, Seattle, WA, USA}  \\
$\circ$ \textit{Department of Statistics, University of British Columbia, BC, Canada} \\ 
$\bigtriangleup$ \textit{Department of Biomedical Engineering, Johns Hopkins University, Baltimore, MD, USA}\\ 
 $\bigtriangledown$ \textit{Department of Computer Science, Johns Hopkins University, Baltimore, MD,  USA} \\
 $\ddagger$ \textit{Department of Biostatistics, University of Washington, Seattle, WA, USA} \\ 
{aneufeld@uw.edu}}
\markboth%
{A. Neufeld and others}
{Inference after latent variable estimation for single-cell RNA sequencing data}

\maketitle

\begin{abstract}
{In the analysis of single-cell RNA sequencing data, researchers often characterize the variation between cells by estimating a latent variable, such as cell type or pseudotime, representing some aspect of the cell's state. They then test each gene for association with the estimated latent variable. 
If the same data are used for both of these steps, then standard methods for computing p-values 
in the second step will fail to achieve 
statistical guarantees such as Type 1 error control. 
Furthermore, approaches such as sample splitting that can be 
applied to solve similar problems in other settings are not applicable
in this context. In this paper, we 
 introduce \emph{count splitting}, a flexible framework that allows us to carry out valid inference in this setting, for virtually any latent variable estimation technique and inference approach, under a Poisson assumption. 
We demonstrate the Type 1 error control and power of count splitting in a simulation study, and apply count splitting to a dataset of pluripotent stem cells differentiating to cardiomyocytes. }
{Poisson, binomial thinning, pseudotime, clustering, selective inference, sample splitting.}
 \end{abstract}

\section{Introduction}
\label{section_introduction}

Techniques for single-cell RNA sequencing (scRNA-seq) allow scientists to measure gene expression of huge numbers of individual cells in parallel. Researchers can then investigate how gene expression varies between cells of different states. In particular, we highlight two common questions that arise in the context of scRNA-seq data: 

\begin{list}{}{}
\item \textbf{Question 1.} \emph{Which genes are differentially expressed along a continuous cellular trajectory?} This trajectory might represent development, activity level of an important pathway, or \emph{pseudotime}, a quantitative measure of biological progression through a process such as cell differentiation \citep{trapnell2014dynamics}. 
\item \textbf{Question 2.}  
\emph{Which genes are differentially expressed between discrete cell types?}  
\end{list}
These two questions are hard to answer because typically the cellular 
trajectory or the cell types are not directly observed, and must be estimated from the data. 
We can unify these two questions, and many others that arise in the analysis of scRNA-seq data, under a latent variable framework.

Suppose that we have mapped the scRNA-seq reads for $n$ cells to $p$ genes (or other functional units of interest). Then, the data matrix $X$ has dimension $n \times p$, and $X_{ij}$ is the number of reads from the $i$th cell that map to the $j$th gene. We assume that $X$ is a realization of a random variable $\bold{X}$, and that the biological variation in $\E[\bold{X}]$ is explained by a set of latent variables $L \in \mathbb{R}^{n \times k}$.  
We wish to know which columns $\bold{X}_j$ of $\bold{X}$ are associated with $L$. As $L$ is unobserved, the following two-step procedure seems natural: \\
\indent \indent \textbf{Step 1: Latent variable estimation.} Use $X$ to compute $\widehat{L}(X)$, an estimate of $L$.  \\
\indent \indent \textbf{Step 2: Differential expression analysis.} For $j = 1,\ldots,p$, test for association between \indent \indent $\bold{X}_j$ and the columns of $\widehat{L}(X)$. \\
In this paper, we refer to the practice of using the same data $X$ to first construct $\hat{L}(X)$ and second to test for association with $\hat{L}(X)$ as ``double dipping". 

\begin{figure}[H]
\centering
\includegraphics[width=0.8\textwidth]{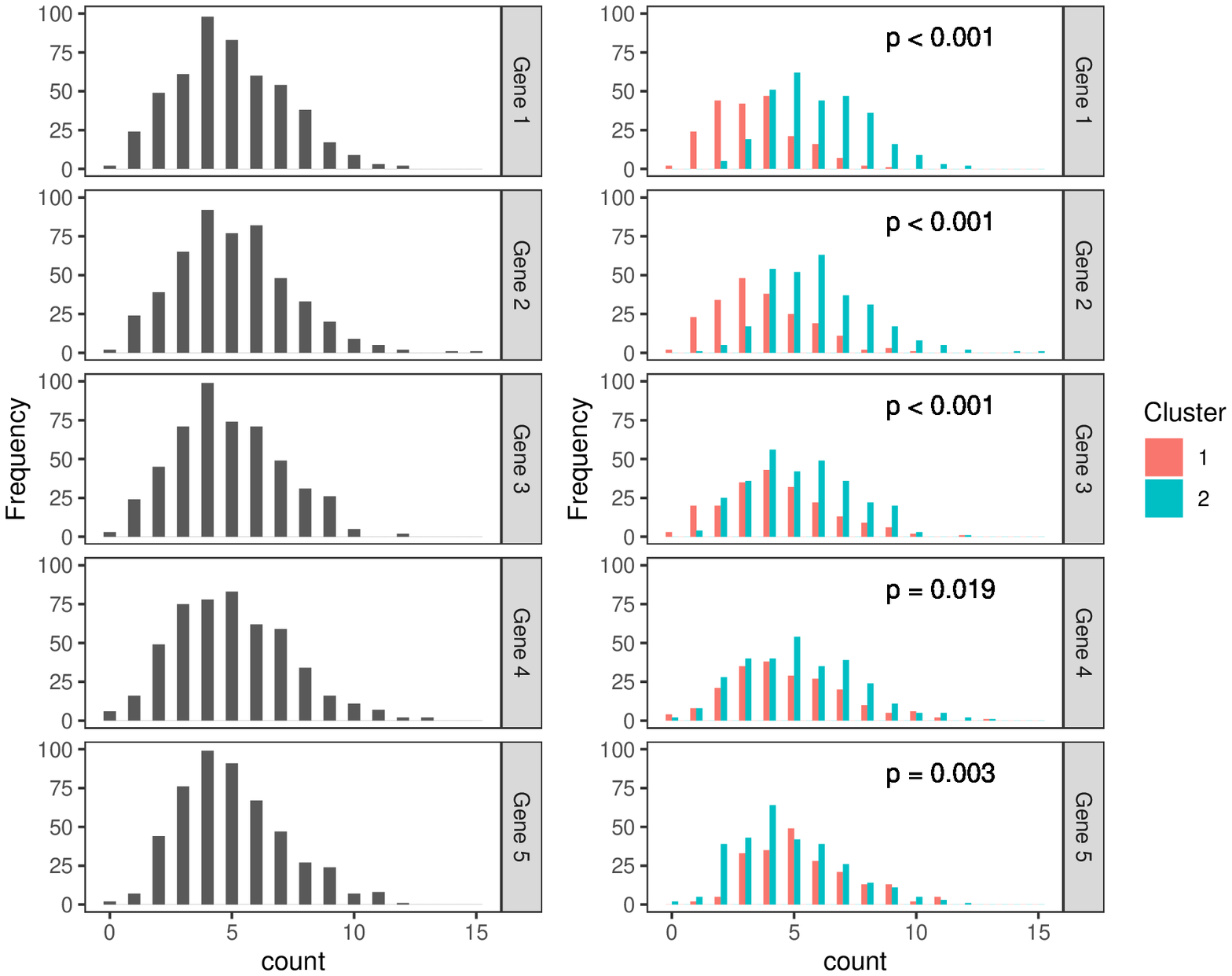}
\caption{\emph{Left:} Distributions of counts for five genes under a model where $\bold{X}_{ij} \sim \mathrm{Poisson}(5)$ for all genes and all cells. \emph{Right:} Distributions of the same counts, colored by estimated cluster, labeled with the Wald p-values from a Poisson GLM. All p-values are small, despite the fact that all null hypotheses hold.}
\label{figure_intro_hists}	
\end{figure}

Why is double dipping a problem? As we will see throughout this paper, if we use standard statistical tests in Step 2, then we will fail to control the Type 1 error rate. 
We now provide intuition for this using a very simple motivating example. We generate $X \in \mathbb{Z}_{\geq 0}^{500 \times 5}$, where ${\bf X}_{ij} \sim \mathrm{Poisson}(5)$ for $i = 1,\ldots,500$ and $j=1,\ldots,5$. The distribution of counts for each of the five genes is shown in the left column of Figure~\ref{figure_intro_hists}. The right panel of Figure~\ref{figure_intro_hists} shows the result of (1) clustering the data by applying k-means with $k=2$,
and then (2) testing for association between each gene and the estimated clusters (using a Wald p-value from a Poisson generalized linear model). Even though all cells are drawn from the same distribution, and thus all tested null hypotheses hold, all five p-values are small. The Type 1 error rate is not controlled, because the Wald test does not account for the fact that the clustering algorithm is designed to maximize the difference between the clusters (see the right panel of Figure~\ref{figure_intro_hists}).

Despite this issue with double dipping, it is common practice. \texttt{Monocle3} and \texttt{Seurat} are popular \texttt{R} packages that each contain functions for (1) estimating latent variables such as clusters or pseudotime, and (2) identifying genes that are differentially expressed across these latent variables. The vignettes for these \texttt{R} packages perform both steps on the same data \citep{monocle3vignette, seuratvignette}. Consequently, the computational pipelines suggested by the package vignettes fail to control the Type 1 error rate. We demonstrate this empirically 
in Appendix A of the supplementary materials.

Though no suitable solution has yet been proposed, the issues associated with this two-step process procedure are well-documented: \cite{lahnemann2020eleven} cite the ``double use of data" in differential expression analysis after clustering as one of the ``grand challenges" in single cell RNA sequencing (Question 1), and \cite{deconinck2021recent} note that the ``circularity" involved in the two-step process of trajectory analysis leads to ``artificially low p-values" for differential expression (Question 2). 

In this paper, we propose \emph{count splitting}, a simple fix for the double dipping problem. This allows us to carry out latent variable estimation and differential expression analysis without double dipping, so as to obtain p-values that
control the Type 1 error for the null hypothesis that a given gene is not associated with an estimated latent variable. 

In Section~\ref{section_background}, we introduce the notation and models that will be used in this paper. In Section~\ref{section_existingmethods}, we carefully examine existing methods for latent variable inference, and explain why they are not adequate in this setting. In Section~\ref{section_method} we introduce our proposed method, and in Sections~\ref{section_simulation} and \ref{sec_realdata} we demonstrate its merits on simulated and real data. 

\section{Models for scRNA-seq data} 
\label{section_background}

Recall that $X_{ij}$ is the number of reads mapping to the $j$th gene in the $i$th cell, and that $X \in \mathbb{Z}_{\geq 0}^{n \times p}$ is a realization from $\bold{X}$. We assume that the entries in $\bold{X}$ are independent, with
\begin{equation}
\label{eq_meanmodel}
\E\left[ \bold{X}_{ij} \right] = \gamma_{i} \Lambda_{ij}, \ \
\log\left(\Lambda_{ij} \right) = \beta_{0j} + \beta_{1j}^T L_i, \ \ i=1,\ldots,n, \ \ j=1,\ldots,p,
\end{equation}
where $\gamma_1,\ldots,\gamma_n$ are cell-specific size factors that reflect technical variation in capture efficiency of the mRNA molecules between cells, the $n \times p$ matrix $\Lambda$ represents biological variation, and the matrix $L \in \mathbb{R}^{n \times k}$ contains the unobserved latent variables. In \eqref{eq_meanmodel}, $\beta_{1j} \in \mathbb{R}^k$ and $L_i$ is the $i$th row of $L$. Throughout this paper, we treat $L$, and thus $\Lambda$, as fixed. 
Estimating $\gamma_1,\ldots,\gamma_n$ can be challenging. As size factor estimation is not the focus of this paper, we assume throughout that the $\gamma_i$ are either known or have been accurately estimated. 

Based on \eqref{eq_meanmodel}, in this paper we carry out the differential expression analysis step of the two-step process introduced in Section~\ref{section_introduction} by fitting a generalized linear model (GLM) with a log link function to predict $X_j$ using $\widehat{L}(X)$,  with the $\gamma_i$ included as offsets. We note that while the notation introduced in the remainder of this section is specific to GLMs, the ideas in this paper can be extended to accommodate alternate methods for differential expression analysis, such as generalized additive models (as in \citealt{van2020trajectory} and \citealt{trapnell2014dynamics}). 

Under the model in \eqref{eq_meanmodel}, the columns of the matrix $\log\left(\Lambda\right)$ are linear combinations of the unobserved columns of $L$. However, they may not be linear combinations of the estimated latent variables, i.e. the columns of $\widehat{L}(X)$. Thus, we need additional notation to describe the population parameters that we target
when we fit a GLM with $\widehat{L}(X)$, rather than $L$, as the predictor. 

Let $p_{\theta}\left( \cdot \right)$ denote the density function of the distribution belonging to the exponential family specified by the GLM with mean $\theta$. For any $Z \in \mathbb{R}^{n \times k}$ and random variable $\bold{X} \in \mathbb{R}^{n \times p}$, we define the population parameters targeted by fitting a GLM with a log link function to predict $X_j$ (drawn from $\bold{X}_j$) using $Z$, with the $\gamma_i$ included as offsets, to be: 
\begin{equation}
\label{eq_betadef}
\Big( \beta_0 \left(Z, \bold{X}_j \right), \beta_1\left(Z, \bold{X}_j\right)  \Big) =  \argmax_{\alpha_0 \in \mathbb{R},\alpha_{1} \in \mathbb{R}^k}  \Bigg( \E_{\bold{X}_{1j},\ldots,\bold{X}_{nj}} \left[\sum_{i=1}^n \log\left( p_{\gamma_i \exp \left( \alpha_0 + \alpha_1^T Z_i \right)}\left( \bold{X}_{ij}\right)\right)\right] \Bigg), 
\end{equation}
where the expectation in \eqref{eq_betadef} is taken over the true joint distribution of $\bold{X}_{1j},\ldots,\bold{X}_{nj}$. We say that the $j$th gene is differentially expressed across a variable $Z$ if $\beta_1\left(Z,\bold{X}_j\right) \neq 0$. If the mean model in \eqref{eq_meanmodel} holds and the distribution of $\bold{X}_{ij}$ belongs to the family specified by the GLM for $i=1,\ldots,n$, then $\beta_1\left( L, \bold{X}_j\right) = \beta_{1j}$ and $\beta_0\left( L, \bold{X}_j\right) = \beta_{0j}$ from \eqref{eq_meanmodel}. More generally, $\beta_{0}(Z, \bold{X}_j)$ and $\beta_{1}(Z, \bold{X}_j)$ are the parameters that make $\prod \limits_{i=1}^n p_{\gamma_i \exp(\alpha_0 + \alpha_1^T Z_i)}(X_{ij})$ closest in Kullback-Leibler (KL) divergence to the true joint distribution for ${\bf X}_{1j}, \ldots, {\bf X}_{nj}$ (see \citealt{wakefield2013bayesian}, Section 2.4.3). 

We denote the coefficient estimates that result from fitting a GLM with a log link function to predict a realized $X_j$ using $Z$, with the $\gamma_i$ included as offsets, as
\begin{equation}
\label{eq_GLM}
\left( \widehat{\beta}_0\left(Z, X_j\right), \widehat{\beta}_1\left(Z, X_j\right)  \right) = \argmax_{\alpha_0 \in \mathbb{R},\alpha_{1} \in \mathbb{R}^k} \sum_{i=1}^n \log \left( p_{\gamma_i \exp \left( \alpha_0 + \alpha_1^T Z_i \right)}\left( X_{ij}\right)\right). 
\end{equation}

For the majority of this paper, as in \cite{wang2018gene}, we assume the model
\begin{equation}
\label{eq-datagen}
\bold{X}_{ij} \overset{\mathrm{ind.}}{\sim} \text{Poisson}(\gamma_i \Lambda_{ij}),
\end{equation} 
and let $p_{\theta}(\cdot)$ be a Poisson density. In Section~\ref{subsec_overdisp}, we discuss the case where ${\bold{X}_{ij}}$ does not follow a Poisson distribution and/or $p_{\theta}(\cdot)$ is not a Poisson density.

\section{Existing methods for latent variable inference}
\label{section_existingmethods}

\subsection{Motivating example}
\label{subsec_motivatingexample}

Throughout Section~\ref{section_existingmethods}, we compare existing methods for differential expression analysis after latent variable estimation on a simple example. We generate $2,000$ realizations of $\bold{X} \in \mathbb{Z}_{\geq 0}^{200 \times 10}$, where $\bold{X}_{ij} \overset{\mathrm{ind.}}{\sim} \text{Poisson}(\Lambda_{ij})$. For $i = 1,\ldots,200$, we let $\Lambda_{ij}=1$ for $j = 1,\ldots,5$ and $\Lambda_{ij}=10$ for $j = 6,\ldots,10$. (This is an example of the model defined in \eqref{eq_meanmodel} and \eqref{eq-datagen} with $\gamma_1 = \ldots = \gamma_n = 1$ and $\beta_{1j} =  0$ for $j = 1,\ldots,10$.)
Under this mechanism, each cell is drawn from the same distribution, and thus there is no true trajectory. Nevertheless, 
for each dataset $X$, we estimate a trajectory using the first principal component of the log-transformed data with a pseudocount of one. We then fit Poisson GLMs (with no size factors) to study differential expression.
Since all columns of $\Lambda$ are constant, $\beta_{1}(Z, \bold{X}_j)=0$ for all $j$ and for any $Z \in \mathbb{R}^n$ (see \eqref{eq_betadef}).
Therefore, any p-value quantifying the association between a gene and an estimated trajectory should follow a $\text{Unif}(0,1)$ distribution. As we will see, most available approaches do not have this behavior.

\subsection{The double dipping method}
\label{subsec_naive}

The \emph{double dipping} method refers to using the same data for latent variable estimation and differential expression analysis, as in the two-step procedure in Section~\ref{section_introduction}, without correcting for this double use. As mentioned in Section~\ref{section_introduction}, variants of this method are used in the vignettes for the popular software packages \texttt{Seurat} \citep{seuratvignette}
and \texttt{Monocle3} \citep{monocle3vignette}. In the context of the motivating example from Section~\ref{subsec_motivatingexample}, this method attempts to do inference on the parameter $ \beta_1\left( \widehat{L}(X), \bold{X}_j\right)$, defined in \eqref{eq_betadef}, by regressing $X_j$ on $\widehat{L}(X)$.  Using notation from Section~\ref{section_background}, the resulting Wald p-values have the form
\begin{equation}
\label{eq_naivep}
\text{Pr}_{H_0:\  \beta_1\left(\widehat{L}(X),\bold{X}_j\right)=0} \left( 
\left| 
\widehat{\beta}_{1}\left(\widehat{L}(X),\bold{X}_j\right) \right|
  \geq 
  \left| 
  \widehat{\beta}_{1}\left(\widehat{L}\left(X\right), X_j\right) \right|
 \right).
\end{equation}
The right-hand side of the inequality in \eqref{eq_naivep} uses the observed data to obtain both the predictor and the response, while the left-hand side does not. Thus, 
$\widehat{\beta}_{1}\left(\widehat{L}(X),X_j\right)$ is drawn from $\widehat{\beta}_{1}\left(\widehat{L}(\bold{X}),\bold{X}_j\right)$, \emph{not}
$\widehat{\beta}_{1}\left(\widehat{L}(X),\bold{X}_j\right)$. To state the issue in a different way, we used the data realization $X$ to construct both the predictor $\hat{L}(X)$ and the response $X_j$ in the GLM, and did not account for this double use in determining the distribution of the test statistic. Therefore, as shown in Figure~\ref{fig_null_qqs}(a), when we compute the p-value in \eqref{eq_naivep} for many realizations of $\bold{X}$, the collection of p-values does not follow a $\text{Unif}(0,1)$ distribution. 

\subsection{Cell splitting}
\label{subsec_sampsplit}

In some settings, it is possible to overcome the issues associated with double dipping
by splitting the observations into a training set and a test set, generating a hypothesis on the training set, and testing it on the test set \citep{cox1975note}. However, in the setting of this paper, splitting the cells in $X$ does not allow us to bypass the issues of double dipping.

Why not? Suppose we estimate the latent variables using the cells in the training set, $X^{\mathrm{train}}$. To test for differential expression using the cells in $X^{\mathrm{test}}$, we need latent variable coordinates for the cells in $X^{\mathrm{test}}$. However, it is not clear how to obtain latent variable coordinates for the cells in $X^\mathrm{test}$ that are only a function of the training set and not the test set (i.e. can be written as $\widehat{L}(X^\mathrm{train})$). In the simple example from Section~\ref{subsec_motivatingexample}, 
after computing the first principal axis of the log-transformed $X^{\mathrm{train}}$ matrix, we could consider projecting the log-transformed $X^{\mathrm{test}}$ onto this axis to obtain coordinates for the cells in $X^{\mathrm{test}}$. Unfortunately, this projection step uses the data in $X^{\mathrm{test}}$, and so the resulting estimated coordinates must be written as $\widehat{L}\left(X^\mathrm{train}, X^\mathrm{test}\right)$. 

If we then fit a Poisson GLM to predict $X_j^\mathrm{test}$ using $\widehat{L}\left(X^\mathrm{train}, X^\mathrm{test}\right)$, 
the Wald p-values from this \emph{cell splitting} procedure have the form
\begin{equation}
\small
\label{eq_cellsplitp}
\text{Pr}_{H_0:\  \beta_1\left(\widehat{L}(X^{\mathrm{train}}, X^{\mathrm{test}}),\bold{X}^\mathrm{test}_j\right)=0} \left( 
\left| 
\widehat{\beta}_{1}\left(\widehat{L}({X}^{\mathrm{train}}, X^{\mathrm{test}}),\bold{X}^\mathrm{test}_j\right) \right|
  \geq 
  \left| 
  \widehat{\beta}_{1}\left(\widehat{L}\left(X^{\mathrm{train}}, X^{\mathrm{test}}\right), X_j^\mathrm{test}\right) \right| \right).
\end{equation}
Unfortunately, \eqref{eq_cellsplitp} suffers from the same issue as \eqref{eq_naivep}: the right-hand side of the inequality uses the same realization ($X^\mathrm{test}$) to construct both the predictor and the response, whereas the left-hand side does not. 
Thus, the p-values from \eqref{eq_cellsplitp} do not follow a $\text{Unif}(0,1)$ distribution, even when the columns of $\Lambda$ are constants. Instead, they are anti-conservative, as shown in Figure~\ref{fig_null_qqs}(a).

\subsection{Gene splitting}
\label{subsec_genesplit}

In the same spirit as cell splitting, we now consider splitting the genes (features), rather than the observations,
to form $X^\mathrm{train}$ and $X^\mathrm{test}$. 
In this setting, $\widehat{L}\left(X^{\mathrm{train}}\right)$ provides coordinates for all cells in $X$, and we can obtain p-values for each gene $X_j$ that is not in $X^{\mathrm{train}}$ by regressing $X_j$ on $\widehat{L}(X^{\mathrm{train}})$. The roles of $X^{\mathrm{train}}$ and $X^{\mathrm{test}}$ can be swapped to obtain p-values for the remaining genes. This \emph{gene splitting} procedure  yields Wald p-values of the form
\begin{align*}
\small
	&\text{Pr}_{H_0:\  \beta_1\left(\widehat{L}(X^{\mathrm{train}}),\bold{X}^\mathrm{test}_j\right)=0} \left( 
\left| 
\widehat{\beta}_{1}\left(\widehat{L}(X^{\mathrm{train}}),\bold{X}^\mathrm{test}_j\right) \right|
  \geq 
  \left| 
  \widehat{\beta}_{1}\left(\widehat{L}\left(X^{\mathrm{train}}\right), X_j^\mathrm{test}\right) \right| \right)  & \text{ if } X_j \in X^{\mathrm{test}}, \\
  &\text{Pr}_{H_0:\  \beta_1\left(\widehat{L}(X^{\mathrm{test}}),\bold{X}^\mathrm{train}_j\right)=0} \left( 
\left| 
\widehat{\beta}_{1}\left(\widehat{L}({X}^\mathrm{test}),\bold{X}^\mathrm{train}_j\right) \right|
  \geq 
  \left| 
  \widehat{\beta}_{1}\left(\widehat{L}\left(X^{\mathrm{test}}\right), X_j^\mathrm{train}\right) \right| \right) & \text{ if } X_j \in X^{\mathrm{train}}.
\end{align*}
As the coefficients on the right-hand side of these inequalities never use the same data to construct the predictor and the response, these p-values will be uniformly distributed over repeated realizations of $\bold{X}$ when the columns of $\Lambda$ are constants. However, they are fundamentally unsatisfactory in scRNA-seq applications where we wish to obtain a p-value for every gene \emph{with respect to the same estimated latent variable.} Thus, gene splitting is not included in Figure~\ref{fig_null_qqs}(a). 

\subsection{Selective inference through conditioning}
\label{subsec_selective}

We next consider taking a \emph{selective inference} approach \citep{lee2016exact, taylor2015statistical} to correct the p-values in \eqref{eq_naivep}. This involves fitting the same regression model as the method that double dips, but replacing~\eqref{eq_naivep} with the conditional probability 
\begin{align}
\label{eq_selectivep}
\text{Pr}_{H_0:\  \beta_1\left(\widehat{L}(X),\bold{X}_j\right)=0} \left( \left| \widehat{\beta}_{1}\left(\widehat{L}(X),\bold{X}_j\right) \right| \geq \left| \widehat{\beta}_{1}\left(\widehat{L}\left(X\right), X_j\right) \right| \mid \widehat{L}\left(\bold{X}\right) = \widehat{L}(X) \right).
\end{align}
The inequality in \eqref{eq_selectivep} is identical to that in \eqref{eq_naivep}, but under the conditioning event,  \eqref{eq_selectivep} can be rewritten as 
\begin{align}
\label{eq_selectivep2}
\text{Pr}_{H_0:\  \beta_1\left(\widehat{L}(X),\bold{X}_j\right)=0} \left( \left| \widehat{\beta}_{1}\left(\widehat{L}(\bold{X}),\bold{X}_j\right) \right| \geq \left| \widehat{\beta}_{1}\left(\widehat{L}\left(X\right), X_j\right) \right| \mid \widehat{L}\left(\bold{X}\right) = \widehat{L}(X) \right), 
\end{align}
such that both the left-hand and right-hand sides of the inequality use the same data to construct the predictor and the response. Thus, over repeated realizations of $\bold{X}$ when the columns of $\Lambda$ are constants, the p-values in \eqref{eq_selectivep} follow a $\text{Unif}(0,1)$ distribution. 

The approach in \eqref{eq_selectivep} is not suitable for the setting of this paper. First, \eqref{eq_selectivep} cannot be computed in practice.  The selective inference literature typically modifies \eqref{eq_selectivep} by conditioning on extra information for computational tractability. Conditioning on extra information does not sacrifice Type 1 error control, and the extra information can be cleverly chosen such that the modified p-value is simple to compute under a multivariate normality assumption. Because scRNA-seq data consists of non-negative integers, a normality assumption is not suitable, and so this approach does not apply. Second, as the conditioning event in \eqref{eq_selectivep} must be explicitly characterized, each choice of $\widehat{L}(\cdot)$ will require its own bespoke strategy. This is 
problematic in the scRNA-seq setting, where many specialized techniques are used for pre-processing, clustering, and trajectory estimation. \citet{zhang2019valid} overcome this challenge in the context of clustering by combining cell splitting (Section~\ref{subsec_sampsplit}) with selective inference. The idea is to condition on the test set labeling event $\left\{ \widehat{L}\left(X^{\mathrm{train}}, \bold{X}^\mathrm{test}\right) = \widehat{L}\left(X^{\mathrm{train}}, X^{\mathrm{test}}\right) \right\}$, which can be characterized provided that the estimated clusters are linearly separable. However, their work requires a normality assumption, and does not extend naturally to the setting of trajectory estimation. 

Selective inference is omitted as a comparison method in Figure~\ref{fig_null_qqs}(a) because we are not aware of a way to compute \eqref{eq_selectivep} for Poisson data, even for our simple choice for $\widehat{L}\left(\cdot\right)$. In Appendix~\ref{appendix_selectivecluster}, we show that the selective inference method of \cite{gao2020selective}, which provides finite-sample valid inference after clustering for a related problem under a normality assumption, 
does not control the Type 1 error rate when applied to log-transformed Poisson data. 

\subsection{Jackstraw}
\label{subsec_jackstraw}

Much of the difficulty with applying a selective inference approach
in the setting of this paper lies in analytically characterizing
the conditioning event in \eqref{eq_selectivep} and \eqref{eq_selectivep2}. An alternative approach is to try to compute a probability similar to \eqref{eq_selectivep2}, but without the conditioning event.
While the distribution of $\widehat{\beta}_{1}\left(\widehat{L}(\bold{X}),\bold{X}_j\right)$ is typically not analytically tractable, its null 
distribution can be approximated via permutation. 
This is the idea behind the \emph{jackstraw} method of \cite{chung2015statistical}, which was originally proposed to test 
for association between the principal components and features of a data matrix, and was later extended to the clustering setting \citep{chung2020statistical}. To make our discussion of jackstraw congruent with the rest of this paper, we instantiate the framework to our latent variable GLM setting, and assume
\begin{equation}
\label{eq_datagen_jackstraw}	
\bold{X}_{ij} \overset{\mathrm{ind.}}{\sim} H(\Lambda_{ij}), 
\ \ \ \ \ 
\log\left(\Lambda_{ij} \right) = \beta_{0j} + \beta_{1j} L_i, \ \ \ \ \beta_{1j}, L_i \in \mathbb{R},
\end{equation}
where $H(\mu)$ is a distribution parameterized by its mean $\mu$. 
Compared to \eqref{eq_meanmodel}, we have omitted the size factors and have assumed that the latent variable is one-dimensional, as these are the assumptions made in the motivating example in Section~\ref{subsec_motivatingexample}. We have additionally assumed that each element of $X$ is independently drawn from a distribution $H(\cdot)$. 

To test whether the $j$th gene is differentially expressed, jackstraw creates datasets $X^\mathrm{permute, b}$ for $b=1,\ldots,B$ by randomly permuting the $j$th column $B$ times. Each dataset gives rise to a GLM coefficient $\widehat{\beta}\left(\widehat{L}(X^{\mathrm{permute,b}}),{X}^\mathrm{permute,b}_j \right)$ and an associated standard error estimate $\widehat{\mathrm{SE}}\left(\widehat{\beta}\left(\widehat{L}(X^{\mathrm{permute,b}}),{X}^\mathrm{permute,b}_j \right)\right)$. The p-value for the $j$th gene is computed as
\begin{equation}
\label{jackstraw_empirical}
\frac{1}{B} \sum_{b=1}^B \bold{1} \left\{  \frac{\left| \widehat{\beta}_1\left(\widehat{L}(X^\mathrm{permute,b}),{X}^\mathrm{permute,b}_j \right)\right|	}{\widehat{\mathrm{SE}}\left(\widehat{\beta}_1\left(\widehat{L}(X^{\mathrm{permute,b}}),{X}^\mathrm{permute,b}_j \right)\right)}   \geq \frac{\left| \widehat{\beta}_1\left(\widehat{L}(X),{X}_j \right) \right|}{\widehat{\mathrm{SE}}\left(\widehat{\beta}_1\left(\widehat{L}(X),{X}_j \right)\right)}  \right\},
\end{equation}
where $\bold{1}\{\cdot\}$ is an indicator function which evaluates to $1$ if the inequality is true and $0$ otherwise. 

Under the null hypothesis that $\beta_{1j} = 0$ in \eqref{eq_datagen_jackstraw}, 
$X^\mathrm{permute,b}$ and  $X$ have the same distribution. Furthermore, both sides of the inequality in \eqref{jackstraw_empirical} use the same data realization to construct the predictor and the response.  This suggests that under the null hypothesis that $\beta_{1j}=0$ in \eqref{eq_datagen_jackstraw}, the distribution of the p-value in \eqref{jackstraw_empirical} will converge to $\text{Unif}(0,1)$ as $B \rightarrow \infty$. This null hypothesis is slightly different than the ones tested in Sections~\ref{subsec_naive}--\ref{subsec_selective}, which involved association between the $j$th gene and an \emph{estimated} latent variable. 

Unfortunately, carrying out jackstraw as described above is computationally infeasible, as testing $H_0: \beta_{1j} = 0$ for $j=1,\ldots,p$
requires  $B \times p$  computations of $\widehat{L}(\cdot)$. To improve computational efficiency, 
\cite{chung2015statistical} suggest randomly choosing a set of $s$ genes $S_b$ to permute to obtain $X^\mathrm{permute,b}$ for $b=1,\ldots,B$. Then, the p-value for the $j$th gene is computed as
\begin{equation}
\label{jackstraw_empirical2}
\frac{1}{B \times s} \sum_{b=1}^B \sum_{q \in S_b}\bold{1} \left\{ \frac{\left|\widehat{\beta}_1\left(\widehat{L}(X^\mathrm{permute,b}),{X}^\mathrm{permute,b}_q \right) \right|}{\widehat{\mathrm{SE}}\left(\widehat{\beta}_1\left(\widehat{L}(X^\mathrm{permute,b}),{X}^\mathrm{permute,b}_q \right)  \right)}	  \geq \frac{\left| \widehat{\beta}_1\left(\widehat{L}(X),{X}_j \right) \right|}{\widehat{\mathrm{SE}}\left(\widehat{\beta}_1\left(\widehat{L}(X),{X}_j \right)  \right)}	  \right\}.
\end{equation} 
Since all $p$ genes are compared to the same reference distribution, only $B$ total computations of $\widehat{L}(\cdot)$ are needed. 
Unfortunately, the p-value in \eqref{jackstraw_empirical2} only follows a $\text{Unif}(0,1)$ distribution if the quantity on the right-hand side of the inequality in \eqref{jackstraw_empirical2} follows the same distribution for  $j=1,\ldots,p$. This does not hold in the simple example from Section~\ref{subsec_motivatingexample}, where the quantity 
has a different distribution for the genes with $\Lambda_{ij}=1$ than those with $\Lambda_{ij}=10$. While the collection of p-values aggregated across all of the genes appears to follow a  $\text{Unif}(0,1)$ distribution (Figure~\ref{fig_null_qqs}(a), left),  some genes have anti-conservative p-values (Figure~\ref{fig_null_qqs}(a), center) and others have overly conservative p-values (Figure~\ref{fig_null_qqs}(a), right).

\subsection{PseudotimeDE}
\label{subsec_pseudode}

\cite{song2021pseudotimede} recently proposed PseudotimeDE to test a gene's association with an estimated trajectory. Here, we present a slight modification of their proposal, which is tailored to our setting but does not change the fundamental properties with respect to the discussion. Implementation details are provided in Appendix~\ref{appendix_fig1help}. For $b=1,\ldots,B$, PseudotimeDE subsamples the cells in $X$ to obtain $X^b$, computes $\widehat{L}(X^{b})$, and then permutes this vector to create $\Pi\left(\widehat{L}(X^{b})\right)$, where $\Pi(\cdot)$ is a permutation operator. It then computes $\widehat{\beta}\left(\Pi\left(\widehat{L}(X^{b})\right), X^b_j \right)$ for each of the $B$ subsamples, for $\hat{\beta}(\cdot,\cdot)$ defined in \eqref{eq_GLM}. The empirical p-value for the $j$th gene is given by
\begin{equation}
\label{eq_pde}	
\frac{1}{B} \sum_{i=1}^B \bold{1}\left\{ 
\left| \widehat{\beta}\left(\Pi\left( \widehat{L}(X^{b})\right), X^b_j \right) \right| \geq \left| \widehat{\beta}\left(\widehat{L}(X), X_j \right) \right|\right\}.
\end{equation}
While it is not entirely clear what null hypothesis PseudotimeDE is designed to test, we can see that there is a problem with the p-value in \eqref{eq_pde}. On the right-hand side of the inequality in \eqref{eq_pde}, there is association between the predictor and the response in the GLM due to the fact that both are generated from the data $X$. On the left-hand side of the inequality in \eqref{eq_pde}, permuting $\widehat{L}(X^b)$ disrupts the association between the predictor and the response. Thus, even in the absence of any signal in the data, 
the quantity on the right-hand side of \eqref{eq_pde} does not have the same distribution as the quantity on the left-hand side of \eqref{eq_pde}. As shown in Figure~\ref{fig_null_qqs}(a), under our simple example where there is no true trajectory, the p-values from \eqref{eq_pde} are anti-conservative.

\section{Count Splitting}
\label{section_method}

\subsection{Method}
\label{subsec_method}

In Sections~\ref{subsec_sampsplit} and ~\ref{subsec_genesplit}, we saw that cell splitting and gene splitting are not suitable options for latent variable inference on scRNA-seq data. Here, we propose \emph{count splitting}, which involves splitting the expression counts themselves, rather than the genes or the cells, to carry out latent variable estimation and differential expression analysis. The algorithm is as follows. 

\begin{algorithm}[Count splitting for latent variable inference]
\label{alg_countsplit}
For a constant $\epsilon$ with $0 < \epsilon < 1$, 
\begin{list}{}{}
\item \textbf{Step 0: Count splitting.} Draw $\bold{X}^{\mathrm{train}}_{ij} \mid \{\bold{X}_{ij} = X_{ij} \} \overset{\mathrm{ind.}}{\sim} \mathrm{Binomial}\left(X_{ij}, \epsilon\right)$, and let $X^{\mathrm{test}} = X-X^{\mathrm{train}}$. 
\item \textbf{Step 1: Latent variable estimation.} Compute $\widehat{L}\left(X^{\mathrm{train}}\right)$.
\item \textbf{Step 2: Differential expression analysis.} For $j = 1,\ldots,p$, 
\begin{enumerate}
\item[(a)] Fit a GLM with a log link to predict $X^{\mathrm{test}}_j$ using $\widehat{L}(X^{\mathrm{train}})$, with the $\gamma_i$ included as offsets. This provides $\widehat{\beta}_1\left( \widehat{L}\left(X^{\mathrm{train}}\right), {X}_j^\mathrm{test}\right)$, an estimate of $\beta_1\left( \widehat{L}\left(X^{\mathrm{train}}\right), \bold{X}_j^\mathrm{test}\right)$. 
\item[(b)] Compute a Wald p-value for $H_0: \beta_1\left( \widehat{L}\left(X^{\mathrm{train}}\right), \bold{X}_j^\mathrm{test}\right)=0$ vs. $H_1: \beta_1\left( \widehat{L}\left(X^{\mathrm{train}}\right), \bold{X}_j^\mathrm{test}\right) \neq 0$, which takes the form
\small
\begin{equation}
\label{eq_countsplitp}	
Pr_{H_0:\  \beta_1\left(\widehat{L}(X^{\mathrm{train}}),\bold{X}^\mathrm{test}_j\right)=0} \left( 
\left| 
\widehat{\beta}_{1}\left(\widehat{L}({X}^\mathrm{train}),\bold{X}^\mathrm{test}_j\right) \right|
  \geq 
  \left| 
  \widehat{\beta}_{1}\left(\widehat{L}\left(X^{\mathrm{train}}\right), X_j^\mathrm{test}\right) \right| \right).
\end{equation}
\normalsize
\end{enumerate} 
\end{list}	
\end{algorithm}

Note that in Step 2(b), we are computing a standard GLM Wald p-value for a regression of $X_j^\mathrm{test}$ onto $\widehat{L}(X^{\mathrm{train}})$. The next result is a well-known property of the Poisson distribution. 

\begin{proposition}[Binomial thinning of Poisson processes (see \citealt{durrett2019probability}, Section 3.7.2)]
\label{prop_independence}
If $\bold{X}_{ij} \sim Poisson(\gamma_i \Lambda_{ij})$, then $\bold{X}_{ij}^\mathrm{train}$ and $\bold{X}_{ij}^\mathrm{test}$, as constructed in Algorithm~\ref{alg_countsplit}, are independent. Furthermore, $\bold{X}_{ij}^\mathrm{train} \sim Poisson(\epsilon \gamma_i \Lambda_{ij})$ and $\bold{X}_{ij}^\mathrm{test} \sim Poisson((1-\epsilon) \gamma_i \Lambda_{ij})$.
\end{proposition}

This means that in \eqref{eq_countsplitp}, under a Poisson model, the predictor and the response are independent on both sides of the inequality. Consequently, the p-value in \eqref{eq_countsplitp} will retain all standard properties of a GLM Wald p-value: e.g. control of the Type 1 error rate for $H_0:\  \beta_1\left(\widehat{L}(X^{\mathrm{train}}),\bold{X}^\mathrm{test}_j\right)=0$ when $n$ is sufficiently large. Furthermore, when $n$ is sufficiently large, we can invert the test in \eqref{eq_countsplitp} to obtain confidence intervals with $100\times(1-\alpha)\%$ coverage for the parameter $\beta_1\left(\widehat{L}(X^{\mathrm{train}}),\bold{X}^\mathrm{test}_j\right)$.

We now consider the parameter $\beta_1\left(\widehat{L}(X^{\mathrm{train}}),\bold{X}^\mathrm{test}_j\right)$.  
Suppose that for any matrix $M$ and scalar $a$, the function $\widehat{L}(\cdot)$ satisfies $\widehat{L}(aM) \propto \widehat{L}(M)$. In this case, Proposition~\ref{prop_independence} says that $\widehat{L}\left( E[\bold{X}^{\mathrm{train}}]\right) = \widehat{L}\left( \epsilon E[\bold{X}] \right) 
\propto \widehat{L}\left(E[\bold{X}]\right)$. 
Furthermore, if
$\log\left( E[\bold{X}_j]\right) = \beta_0 + \beta_1^T L_i$, 	
then 
$\log\left(E[\bold{X}^\mathrm{test}_j]\right) = \log(1-\epsilon) + \beta_0 + \beta_1^T L_i.$ 
Therefore, $\beta_1\left(\widehat{L}(X^{\mathrm{train}}),\bold{X}^\mathrm{test}_j\right)$ is closely related to $\beta_1\left(\widehat{L}(X),\bold{X}_j\right)$, the parameter (unsuccessfully) targeted by the double dipping method (Section~\ref{subsec_naive}). 
\begin{remark}
The key insight behind Algorithm~\ref{alg_countsplit} is that, under a Poisson assumption, a test for association between $\widehat{L}\left(X^\mathrm{train}\right)$ and $\bold{X}_j^\mathrm{test}$ that uses $\widehat{L}\left(X^\mathrm{train}\right)$ and $X_j^\mathrm{test}$ will inherit standard statistical guarantees, despite the fact that $\widehat{L}\left(X^\mathrm{train}\right)$ and $X_j^\mathrm{test}$ are functions of the same data. This insight did not rely on the use of a GLM or a Wald test in Step 2 of Algorithm~\ref{alg_countsplit}. Thus, other approaches could be used to quantify this association. 
\end{remark}
Figure~\ref{fig_null_qqs}(a) shows that count splitting with $\epsilon=0.5$ produces uniformly distributed p-values for all genes in the example described in Section~\ref{subsec_motivatingexample}, using a Poisson GLM in Step 2 of Algorithm~\ref{alg_countsplit}. We explore more values of $\epsilon$ and more complicated scenarios in Section~\ref{section_simulation}.

Count splitting is a special case of ``data fission", proposed in a preprint by \cite{leiner2022data} while this paper was in preparation. While the data fission framework is broad enough to encompass this latent variable setting, \cite{leiner2022data} focus on comparing data fission to sample splitting in supervised settings where the latter is an option. In unsupervised settings, where sample splitting is not an option (as seen in Section~\ref{subsec_sampsplit}), 
ideas similar to count splitting have been applied by \cite{batson2019molecular} and \cite{chen2021estimating} for tasks such as evaluating the goodness-of-fit of a low-rank approximation of a matrix. We elaborate on connections to \cite{batson2019molecular} in Appendix~\ref{appendix_mcv} of the supplementary materials. 
Finally, \cite{gerard2020data} suggest applying binomial thinning to real scRNA-seq datasets to generate synthetic datasets to use for comparing and evaluating scRNA-seq software packages and methods, and this approach is used in the supplement of \cite{sarkar2021separating} to compare various scRNA-seq models. 

\subsection{What if the data are not Poisson?}
\label{subsec_overdisp}

The independence between $\bold{X}_{ij}^\mathrm{train}$ and $\bold{X}_{ij}^\mathrm{test}$ in Proposition~\ref{prop_independence} holds if and only if $\bold{X}_{ij}$ has a Poisson distribution \citep{kimeldorf1981simultaneous}. 
Thus, if we apply Algorithm~\ref{alg_countsplit} to data that are not Poisson, there will be dependence between the predictor and the response on the right-hand side of the inequality in \eqref{eq_countsplitp}, and the p-value in \eqref{eq_countsplitp} will not be uniformly distributed under $H_0$. 

\cite{wang2018gene} argue that the Poisson model is sufficient to model scRNA-seq data. \cite{townes2019feature} advocate modeling the counts for each cell with a multinomial distribution. When the number of genes is large, the elements of the multinomial can be well-approximated by independent Poisson distributions, and indeed the authors rely on this approximation for computational reasons. 
Similarly, \cite{batson2019molecular} assume that $\bold{X}_{ij} \sim \mathrm{Binomial}(\Omega_{ij}, p_i)$, but note that typically $\Omega_{ij}$ is large and $p_i$ is small, so that a Poisson approximation applies. 

\cite{sarkar2021separating} advocate pairing a Poisson \emph{measurement} model for scRNA-seq data with a separate \emph{expression} model to account for overdispersion compared to the Poisson model. For example, a Gamma expression model leads to
\begin{equation}
\label{eq_negbinmech}
\bold{X}_{ij} \mid \{ \boldsymbol{\tau}_{ij} = \tau_{ij} \} \sim \text{Poisson}\left(\Lambda_{ij}{\tau}_{ij}\right), \ \ \ \boldsymbol{\tau}_{ij} \sim \text{Gamma}(b_j,b_j),
\end{equation}
where we have omitted size factors for simplicity. This induces a negative binomial marginal distribution on $\bold{X}_{ij}$, where $\E[\bold{X}_{ij}] = \Lambda_{ij}$ and $\Var\left(\bold{X}_{ij}\right) = \Lambda_{ij} + \frac{\Lambda_{ij}^2}{b_j}$. If $\bold{X}_{ij}$ is drawn from \eqref{eq_negbinmech} and Algorithm~\ref{alg_countsplit} is applied, then $\bold{X}_{ij}^\mathrm{train}$ and $\bold{X}_{ij}^\mathrm{test}$ each follow negative binomial distributions, with $\E[\bold{X}_{ij}^\mathrm{train}] = \epsilon \E[\bold{X}_{ij}]$ and $\E[\bold{X}^\mathrm{test}] = (1-\epsilon) \E[\bold{X}]$ \citep{harremoes2010thinning}. Our next result, proven in Appendix~\ref{appendix_negbincor}, quantifies the correlation between $\bold{X}_{ij}^\mathrm{train}$ and $\bold{X}_{ij}^\mathrm{test}$ in this setting. 

\begin{proposition}
\label{prop_negbincor}
Suppose that $\bold{X}_{ij}$ follows a negative binomial distribution with expected value $\Lambda_{ij}$ and variance $\Lambda_{ij} +\frac{\Lambda_{ij}^2}{b_j}$. If we perform Step 0 of Algorithm~\ref{alg_countsplit}, then  
\begin{equation}
\label{eq_correlation}
\mathrm{Cor}\left(\bold{X}^{\mathrm{train}}_{ij}, \bold{X}^{\mathrm{test}}_{ij}\right) = \frac{\sqrt{\epsilon(1-\epsilon)}}{\sqrt{\epsilon(1-\epsilon)+\frac{b_j^2}{\Lambda_{ij}^2}+\frac{b_j}{\Lambda_{ij}}}}. 
\end{equation}
\end{proposition}

To investigate the performance of count splitting under overdispersion, we generate datasets under \eqref{eq_negbinmech} with $n=200$ and $p=10$. For each dataset, $\Lambda_{ij} = \Lambda = 5$ for $i=1,\ldots,n$ and $j=1,\ldots,p$, and $b_j=b$ for $j=1,\ldots,p$, so that every element of $X$ is drawn from the same distribution. We generate 500 datasets for each value of $b \in \{50,10,5,0.5\}$, so that $\frac{\Lambda}{b} \in \{0.1, 0.5, 1, 10\}$. Figure~\ref{fig_null_qqs}(b) displays the Wald p-values that result from running Algorithm~\ref{alg_countsplit} with a negative binomial GLM in Step 2, for all genes across all of the simulated datasets.

The denominator of \eqref{eq_correlation} in Proposition~\ref{prop_negbincor} shows that $\frac{\Lambda}{b}$ determines the extent of correlation between $\bold{X}^{\mathrm{train}}$ and $\bold{X}^{\mathrm{test}}$ in this setting. As shown in Figure~\ref{fig_null_qqs}(b), when $\frac{\Lambda}{b}$ is small, count splitting produces approximately uniformly distributed p-values, and thus comes very close to controlling the Type 1 error rate. As $\frac{\Lambda}{b}$ grows, the performance of count splitting approaches that of the double dipping method discussed in Section~\ref{subsec_naive}. Thus, count splitting tends to outperform the double dipping method, and in the case of extremely high overdispersion will be \emph{no worse than} the double dipping method. For the real scRNA-seq dataset considered in Section~\ref{sec_realdata}, we show in Appendix~\ref{appendix_realdata} that the majority of the estimated values of $\frac{\Lambda_{ij}}{b_j}$ are less than $1$. 

\subsection{Choosing the tuning parameter $\epsilon$}
\label{subsec_epsilon}

The parameter $\epsilon$ in Algorithm~\ref{alg_countsplit} governs a tradeoff between the information available for estimating $L$  and the information available for carrying out inference. Proposition~\ref{prop_cor_poisson}, proven in Appendix~\ref{proof_cor_poisson}, formalizes the intuition that $X^{\mathrm{train}}$ will look more similar to $X$ when $\epsilon$ is large. 

\begin{proposition}
\label{prop_cor_poisson}	
If $\bold{X}_{ij} \sim \mathrm{Poisson}(\gamma_i \Lambda_{ij})$, then
$\mathrm{Cor}(\bold{X}_{ij}, \bold{X}_{ij}^\mathrm{train}) = \sqrt{\epsilon}.$
\end{proposition}
Thus, as $\epsilon$ decreases, we expect $\widehat{L}(X^{\mathrm{train}})$ and $\widehat{L}(X)$ to look less similar. This is a drawback, as scientists would ideally like to estimate $L$ using \emph{all} of the data. 
However, as $\epsilon$ increases, the power to reject false null hypotheses in Step 2(b) of Algorithm~\ref{alg_countsplit} decreases. Proposition~\ref{prop_power}, proven in Appendix~\ref{power_proof},
quantifies this loss of power. 
\begin{proposition}
\label{prop_power}
Let $
\bold{X}_{ij} \overset{\mathrm{ind.}}{\sim} \mathrm{Poisson}\left( \gamma_i \exp(\beta_{0j} +\beta_{1j} L_i )\right)
$. 	Then 
$
\text{Var}\left(\widehat{\beta}_1(L, \bold{X}^\mathrm{test}_j)\right) \approx \frac{1}{1-\epsilon} \text{Var}\left(\widehat{\beta}_1(L, \bold{X}_j)\right).
$
\end{proposition}
In the ideal setting where $\widehat{L}(X^{\mathrm{train}})=L$ and $L \in \mathbb{R}^{n \times 1}$ for simplicity, using $\bold{X}_j^\mathrm{test}$ rather than $\bold{X}_j$ as the response inflates the variance of the estimated coefficient  by a factor of $\frac{1}{1-\epsilon}$. Thus, when $\epsilon$ is large, Step 2(b) of Algorithm \ref{alg_countsplit} has lower power. In practice, we recommend setting $\epsilon=0.5$ to balance the dual goals of latent variable estimation and downstream inference. 

\section{Simulation study}
\label{section_simulation}

\subsection{Data generating mechanism}
\label{subsec_setup}

We generate data from \eqref{eq_meanmodel} and \eqref{eq-datagen} with $n=2700$ and $p=2000$. We generate the size factors $\gamma_i \overset{\mathrm{ind.}}{\sim} \text{Gamma}\left(10, 10\right)$ and treat them as known. Throughout this section, whenever we perform count splitting, we fit a Poisson GLM in Step 2 of Algorithm~\ref{alg_countsplit} and report Wald p-values.

In this section, we investigate the performance of count splitting under two data-generating mechanisms: one generates a true continuous trajectory and the other generates true clusters. To generate data with an underlying trajectory, we set $L = \left(I_n - \frac{1}{n} 11^T\right) Z$, where $Z_i \overset{\mathrm{ind.}}{\sim} N(0,1)$. Under \eqref{eq_meanmodel}, $L$ is the first principal component of the matrix $\log(\Lambda)$. To estimate $L$, we take the first principal component of the matrix $\log(\text{diag}(\gamma)^{-1} X + 11^T)$. To generate data with underlying cell types, we let $L_i \overset{\mathrm{ind.}}{\sim}\text{Bernoulli}(0.5)$ in \eqref{eq_meanmodel}, indicating membership in one of two cell types. We estimate $L$ by running $k$-means with $k=2$ on the matrix $\log\left (\text{diag}(\gamma)^{-1} X + 11^T\right)$. (In each case, we have used a pseudocount of one to avoid taking the log of zero.) 

In our primary simulation setting, the value $\beta_{0j}$ for each gene is randomly chosen to be either $\log(3)$ or $\log(25)$ with equal probability, such that the data includes a mix of low-intercept and high-intercept genes. For each dataset, we let $\beta_{1j}=0$ for $90\%$ of the $p$ genes. Under \eqref{eq_meanmodel}, 
$\Lambda_{1j} = \ldots = \Lambda_{nj}$ for these genes and thus $\beta_1\left(Z, \bold{X}_j \right)=0$ for any estimated latent variable $Z$. Thus, we refer to these as the null genes. The remaining 10\% of the genes have the same non-zero value of $\beta_{1j}$; these are the differentially-expressed genes. For each latent variable setting, we generate $100$ datasets for each of 15 equally-spaced values of $\beta_{1j}$ in $[0.18, 3]$.

\subsection{Type 1 error results}
\label{subsec_type1}

Figure~\ref{fig_simres}(a) shows that count splitting controls the Type 1 error rate for a range of $\epsilon$ values for the 90\% of genes for which $\beta_{1j}=0$ in datasets that are generated as specified in Section~\ref{subsec_setup}.

\subsection{Quality of estimate of $L$}
\label{subsec_detection}

As mentioned in Section~\ref{subsec_epsilon}, smaller values of $\epsilon$ in count splitting compromise our ability to accurately estimate the unobserved latent variable $L$. 
To quantify the quality of our estimate of $L$, in the trajectory estimation case we compute the absolute value of the correlation between $L$ and $\widehat{L}\left(X^{\mathrm{train}}\right)$, and in the clustering setting we compute the adjusted Rand index \citep{hubert1985comparing}
between $L$ and $\widehat{L}\left(X^{\mathrm{train}}\right)$. The results are shown in Figure~\ref{fig_simres}(b). Here, we consider three settings for the value of $\beta_{0j}$: (i) each gene is equally likely to have $\beta_{0j}=\log(3)$ or $\beta_{0j}=\log(25)$ (as in Section~\ref{subsec_type1}); (ii) all genes have  $\beta_{0j}=\log(3)$ (``100\% low-intercept" setting); and (iii) all genes have  $\beta_{0j}=\log(25)$ (``100\% high-intercept" setting). The 100\% low-intercept setting represents a case where the sequencing was less deep, and thus the data more sparse. 

After taking a log transformation, the genes with $\beta_{0j}=\log(3)$ have higher variance than those with $\beta_{0j}=\log(25)$. Thus, there is more noise in the ``100\% low-intercept" setting than in the ``100\% high-intercept" setting. As a result, for a given value of $\epsilon$, the ``100\% low-intercept" setting results in the lowest-quality estimate of $L$. Estimation of $L$ is particularly poor in the ``100\% low-intercept" setting when $\epsilon$ is small, since then $X^{\mathrm{train}}$ contains many zero counts. This suggests that count splitting, especially with small values of $\epsilon$, is less effective on shallow sequencing data.

\subsection{Power results}
\label{subsec_power}

For each dataset and each gene, we compute the true parameter $\beta_{1}\left(\widehat{L}\left(X^{\mathrm{train}}\right), \bold{X}_j^\mathrm{test}\right)$ defined in \eqref{eq_betadef}. For a Poisson GLM (i.e. $p_\theta(\cdot)$ in \eqref{eq_betadef} is the density of a $\mathrm{Poisson}(\theta)$ distribution), it is straightforward to show that we can compute
$\beta_{1}\left(\widehat{L}\left(X^{\mathrm{train}}\right), \bold{X}_j^\mathrm{test}\right)$ by fitting a Poisson GLM with a log link to predict $E\left[ \bold{X}_{j} \right]$ using $\widehat{L}\left(X^\mathrm{train}\right)$, with the $\gamma_i$ included as offsets. 

\begin{figure}[H]
\centering
\includegraphics[width=\textwidth]{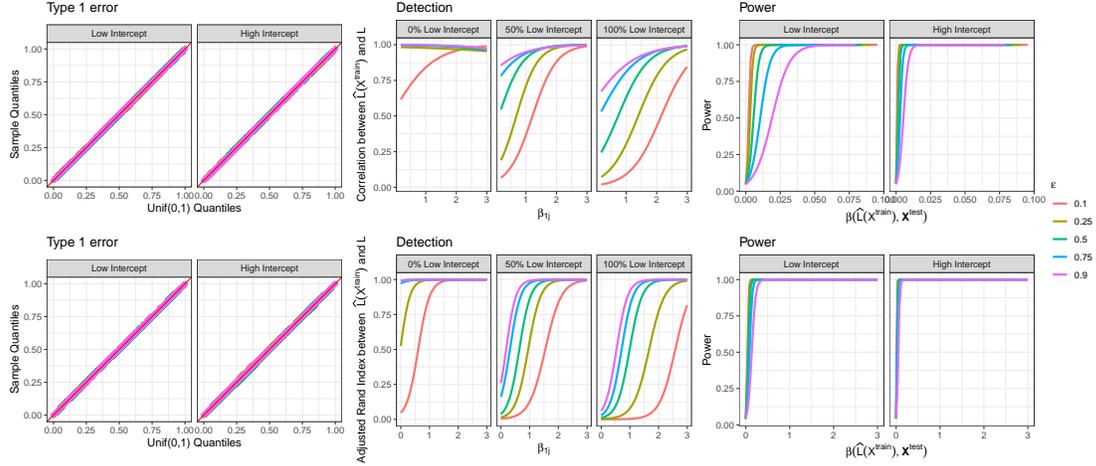}
\caption{
\emph{(a)} Uniform QQ plots of GLM p-values for all null genes from the simulations described in Section~\ref{subsec_setup}. \emph{(b)} Quality of estimate of $L$, as defined in Section~\ref{subsec_detection}, as a function of $\beta_{1j}$ and the percent of low-intercept genes in the dataset.
\emph{(c)} The proportion of null hypotheses that are rejected, aggregated across all non-null genes for all datasets generated with 50\% low-intercept and 50\% high-intercept genes. The parameter $\beta_1\left(\hat{L}(X^\mathrm{train}), \bold{X}_j^\mathrm{test}\right)$ is defined in \eqref{eq_betadef}.
 }
\label{fig_simres}	
\end{figure}

Figure~\ref{fig_simres}(c) shows that our ability to reject the null hypothesis depends on this true parameter, as well as on the value of $\epsilon$. These results are shown in the setting where 50\% of the genes have intercept $\log(3)$ and the other 50\% have intercept $\log(25)$. The impact of $\epsilon$ on power is more apparent for genes with smaller intercepts, as these genes have even less information left over for inference when $(1-\epsilon)$ is small. This result again suggests that count splitting will work best on deeply sequenced data where expression counts are less sparse.

As suggested by Section~\ref{subsec_epsilon}, Figures~\ref{fig_simres}(a) and \ref{fig_simres}(c) show a tradeoff in choosing $\epsilon$: a larger value of $\epsilon$ improves latent variable estimation, but yields lower power for differential expression analysis. In practice, we recommend choosing $\epsilon=0.5$. 

\subsection{Coverage results}
\label{subsec_coverage}

For each dataset and each gene, we compute a 95\% Wald confidence interval for the slope parameter in the GLM. As shown in Table~\ref{tab_coverage}, these intervals achieve nominal coverage for the target parameter $\beta_{1}\left(\widehat{L}\left(X^{\mathrm{train}}\right), \bold{X}_j^\mathrm{test}\right)$, defined in \eqref{eq_betadef} and computed as in Section~\ref{subsec_power}.

\section{Application to cardiomyocyte differentiation data}
\label{sec_realdata}

\cite{elorbany2022single} collect single-cell RNA-sequencing data at seven unique time points (over 15 days) in 19 human cell lines. The cells began as induced pluripotent stem cells (IPSCs) on day 0, and over the course of 15 days they differentiated along a bifurcating trajectory into either cardiomyocytes (CMs) or cardiac fibroblasts (CFs). We are interested in studying  genes that are differentially expressed along the trajectory from IPSC to CM. Throughout this analysis, we ignore the true temporal information (the known day of collection). 

Starting with the entire dataset $X$, we first perform count splitting with $\epsilon=0.5$ to obtain $X^\mathrm{train}$ and $X^\mathrm{test}$. We then perform the lineage estimation task from \cite{elorbany2022single} on $\mathrm{X}^\mathrm{train}$ to come up with a subset of 10,000 cells estimated to lie on the IPSC to CM trajectory. We retain only these 10,000 cells for further analysis. In what follows, to facilitate comparison between the double dipping method and our count splitting approach, we will use this same subset of 10,000 cells for all methods, even though in practice the double dipping method would have chosen the lineage subset based on \emph{all} of the data, rather than on $X^\mathrm{train}$ alone. 

Next, we estimate a continuous differentiation trajectory using 
the \texttt{orderCells()} function from the \texttt{Monocle3} package in \texttt{R} \citep{monocle3vignette}. Details are given in Appendix~\ref{appendix_mymonocle}. To estimate the size factors $\gamma_1,\ldots,\gamma_n$ in \eqref{eq-datagen}, we let $\hat{\gamma}_i(X)$ be the normalized row sums of the expression matrix $X$, as is the default in the \texttt{Monocle3} package. Throughout this section, for $j=1,\ldots,p$, we compare three methods:
\begin{list}{}{}
\item \hspace{-5mm} \emph{Full double dipping:} Fit a Poisson GLM of $X_j$ on $\widehat{L}(X)$ with $\hat{\gamma}_i(X)$ included as offsets. 
\item \hspace{-5mm} \emph{Count splitting:} Fit a Poisson GLM of $X^\mathrm{test}_j$ on $\widehat{L}(X^\mathrm{train})$ with $\hat{\gamma}_i(X^\mathrm{train})$ included as offsets. 
\item \hspace{-5mm} \emph{Test double dipping:} Fit a Poisson GLM of $X^\mathrm{test}_j$ on $\widehat{L}(X^\mathrm{test})$ with $\hat{\gamma}_i(X^\mathrm{test})$ included as offsets.  
\end{list}
For each method, we report the Wald p-values for the slope coefficients. The test double dipping method is included to facilitate understanding of the results from the other two methods. 
We show in Appendix~\ref{appendix_realdata} that the estimated overdispersion parameters are relatively small for this dataset, justifying the use of 
count splitting and Poisson GLMs in each of the methods above. In conducting count splitting, we ensure that none of the pre-processing steps required to compute $\widehat{L}(X^\mathrm{train})$ make use of $X^\mathrm{test}$. 

We first analyze all $10,000$ cells. There is a true differentiation trajectory in this dataset: cells were sampled at at seven different time points of a directed differentiation protocol, and changes in gene expression as well as phenotype that are characteristic of this differentiation process were observed \citep{elorbany2022single}. We focus on a subset of $p=2,500$ high-variance genes that were selected from more than $32,000$ genes that were expressed in the raw data (see Appendix~\ref{appendix_mymonocle} for details). The left panel of Figure~\ref{fig_realdata} displays the count splitting p-values for differential expression of these 2,500 genes against the double dipped p-values. 

We first note the general agreement between the three methods: genes that have small p-values with one method generally have small p-values with all the methods. Thus, count splitting tends to identify the same differentially expressed genes as the methods that double dip when there is a true trajectory in the data. 
That said, we do notice that the full double dipping method tends to give smaller p-values than count splitting. There are two possible reasons for this: (i) the double dipped p-values may be artificially small due to the double dipping, or
(ii) it might be due to the fact that the full double dipping method uses twice as much data. In this setting, it seems that (ii) is the correct explanation: the test double dipping method (which uses the same amount of data as count splitting both to estimate $L$ and to test the null hypothesis) does not yield smaller p-values than count splitting. It seems that there is enough true signal in this data that most of the genes identified by the double dipped methods as differentially expressed are truly differentially expressed, rather than false positives attributable to double dipping. 

Next, we subset the 10,000 cells to only include the $2,303$ cells that were measured at Day $0$ of the experiment, before differentiation had begun.  Then, despite knowing that these cells should be largely homogeneous, we apply $\widehat{L}(\cdot)$ to estimate pseudotime. We note that the $\widehat{L}(\cdot)$ function described in Appendix~\ref{appendix_mymonocle} controls for cell cycle-related variation in gene expression so that this signal cannot be mistaken for pseudotime. We suspect that in this setting, any association seen between pseudotime and the genes is due to overfitting or random noise. We select a new set of $2,500$ high variance genes using this subset of cells. The right panel of Figure~\ref{fig_realdata} shows uniform QQ plots of the p-values for these $2,500$ genes from each of the three methods. In the absence of real differentiation signal, the p-values should follow a uniform distribution. We see that count splitting controls the Type 1 error rate, while the methods that double dip do not. Unlike the previous example that had a very strong true differentiation signal, even test double dipping yields false positives.  Without assessment, the level of true signal in any dataset cannot be assumed to outweigh the effects of double dipping.

In summary, count splitting detects differentially expressed genes when there is true signal in the data, and protects against Type 1 errors when there is no true signal in the data.

\section{Discussion}
\label{section_discussion}

Under a Poisson assumption, count splitting provides a flexible framework for carrying out valid inference after latent variable estimation that can be applied to virtually any latent variable estimation method and inference technique. This has important applications in the growing field of pseudotime or trajectory analysis, as well as in cell type analysis. 

We expect count splitting to be useful in situations other than those considered in this paper. For example, as explored in Appendix~\ref{appendix_selectivecluster}, count splitting can be used to test the overall difference in means between two estimated clusters, as considered by \cite{gao2020selective} and \cite{chen2022selective}. 
As suggested in related work by \cite{batson2019molecular}, \cite{chen2021estimating}, and \cite{gerard2020data}, count splitting could be used for model selection tasks such as choosing  how many dimensions to keep when reducing the dimension of Poisson scRNA-seq data.
Finally, count splitting may lead to power improvements over sample splitting for Poisson data on tasks where the latter is an option \citep{leiner2022data}. 

In this paper, we assume that after accounting for heterogenous expected expression across cells and genes, the scRNA-seq data follows a Poisson distribution. Some authors have argued that scRNA-seq data is overdispersed. As discussed in Section~\ref{subsec_overdisp}, count splitting fails to provide independent training and testing sets when applied to negative binomial data. \cite{leiner2022data} suggest that inference can be carried out in this setting by working with the conditional distribution of $X^\mathrm{test} \mid X^{\mathrm{train}}$. Unfortunately, this conditional distribution does not lend itself to inference on parameters of interest. In future work, we will consider splitting algorithms for overdispersed count distributions. 

Code for reproducing the simulations and real data analysis in this paper is available at \texttt{github.com/anna-neufeld/countsplit\_paper}. An \texttt{R} package with tutorials is available at \\\texttt{anna-neufeld.github.io/countsplit}. 

\section{Acknowledgements}

Anna Neufeld and Daniela Witten were funded by the Simons Foundation (Simons Investigator Award in Mathematical Modeling of Living Systems). Lucy Gao was supported by the Natural Sciences and Engineering Research Council of Canada (Discovery Grants). Alexis Battle was funded by NIGMS MIRA - 1R35GM139580.

\bibliographystyle{biorefs}
\bibliography{latent.bib}

\newpage 

\begin{figure}[H]
\includegraphics[width=\textwidth]{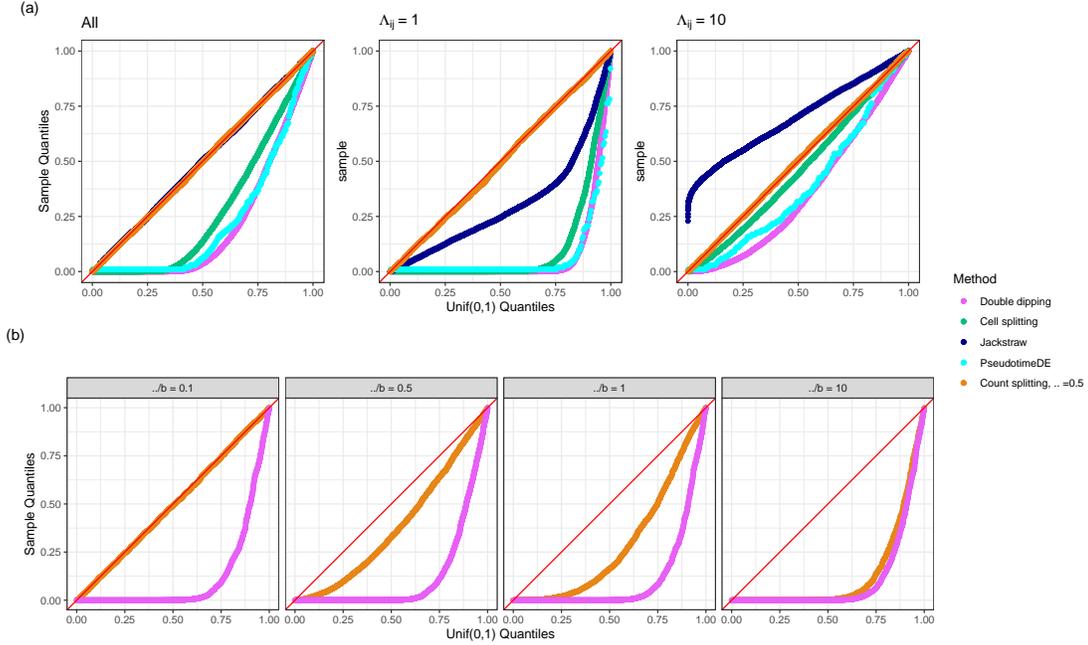}
\caption{Uniform QQ-plots of p-values under the null. 
\emph{(a):} The data are Poisson. P-values are displayed for the $p=10$ genes in each of 2000 datasets generated as described in Section~\ref{subsec_motivatingexample}. The left-hand panel shows all of the genes aggregated, whereas the center and right panels break the results down by the value of $\Lambda_{ij}$.
\emph{(b):} The data are negative binomial. P-values are displayed for the $p=10$ genes in each dataset for the simulation described in Section~\ref{subsec_overdisp}. Results are broken down by the magnitude of $\frac{\Lambda}{b}$. As $\frac{\Lambda}{b}$ increases, the correlation between $\bold{X}^{\mathrm{train}}$ and $\bold{X}^{\mathrm{test}}$ increases, and the performance of count splitting approaches that of the double dipping method. 
}
\label{fig_null_qqs}
\end{figure}

\begin{table}[H]
\centering
\begin{tabular}{c ccc ccc}
&& \multicolumn{2}{c}{Trajectory Estimation} && \multicolumn{2}{c}{Clustering}  \\[5pt]
\hline
 && Low Intercept & High Intercept && Low Intercept & High Intercept  \\[5pt]
 \hline
Null genes &&  0.950 & 0.950 && 0.949  &  0.950 \\
Non-null genes && 0.955 & 0.956 && 0.949 &  0.950  \\
\end{tabular}
\caption{The proportion of 95\% confidence intervals that contain $\beta_{1}\left(\widehat{L}\left(X^{\mathrm{train}}\right), \bold{X}_j^\mathrm{test}\right)$, aggregated across genes and across values of $\epsilon$. }
\label{tab_coverage}	
\end{table}

\begin{figure}[H]
\includegraphics[width=\textwidth]{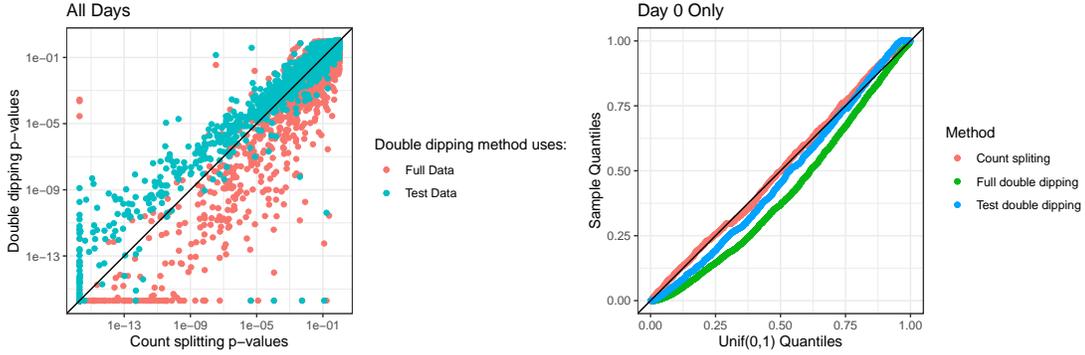}	
\caption{\emph{Left:} Count splitting p-values vs. double dipped p-values for $2,500$ genes, on a log scale, when all $10,000$ cells are used in the analysis. \emph{Right: } Uniform QQ plot of p-values for $2,500$ genes obtained from each method when only the Day 0 cells are used.}
\label{fig_realdata}
\end{figure}


\appendix

\section{Illustrating the issues with the standard Seurat and Monocle3 pipelines}
\label{appendix_seurat_monocle}

In this appendix, we elaborate on our claim that the popular \texttt{R} packages \texttt{Seurat} \citep{seuratpackage} and \texttt{Monocle3} \citep{monocle3package} promote use of the problematic two step procedure that we introduced in Section~\ref{section_introduction} and that we refer to as ``double dipping". The individual functions in both packages are sound. However, because the pipelines suggested in the package vignettes make use of the same data for both latent variable estimation and inference (``double dipping"), they lead to artificially small p-values that fail to control the Type 1 error rate. 

The \texttt{Seurat} ``guided clustering tutorial"  \citep{seuratvignette} suggests that, after applying numerous preprocessing steps, the function \texttt{FindClusters()} should be applied to the data to estimate clusters, and then subsequently the function \texttt{FindMarkers()} should be applied to the same data to test for differential expression across these estimated clusters. The \texttt{FindMarkers()} function returns p-values from standard statistical tests (Wilcoxon rank sum tests by default, or other tests if different arguments are supplied) that do not account for the fact that the clusters were estimated from the data. 

Similarly, the \texttt{Monocle3} tutorial contains a section called ``finding genes that change as a function of pseudotime" \citep{monocle3vignette}. After preprocessing and dimension reduction, the functions \texttt{learn\_graph()} and \texttt{order\_cells()} are applied to the data to estimate pseudotime. Subsequently, the function \texttt{graph\_test()} is applied to the same data. This function applies a standard Moran's I test to test for spatial dependence of gene expression across pseudotime space for each gene, which does not correct for the fact that the pseudotime space was estimated from the data. 

To demonstrate the deficiencies with the above pipelines, we perform a simple simulation study. We generate 500 datasets $X \in \mathbb{Z}_{\geq 0}^{500 \times 200}$ where $\bold{X}_{ij} \sim \mathrm{Poisson}(5)$. Under this data generating mechanism, all cells are drawn from the same distribution and so no genes are differentially expressed across any latent variables. We then carry out the following two methods for each dataset $X$. 
\begin{itemize}
\item \textbf{Seurat (double dipping, as is done in the Seurat tutorial): }	We convert $X$ into a \texttt{Seurat} object. We run the required preprocessing steps, and then we apply the \texttt{FindClusters()} function with resolution = 1 to the object. We then run \texttt{FindMarkers()} on the object to test for differential expression between the first and second estimated clusters. We save the Wilcoxon rank-sum p-values for each gene returned by the \texttt{FindMarkers()} function. 
\item \textbf{Seurat (count splitting): } We first perform count splitting on the data to obtain $X^{\mathrm{train}}$ and $X^{\mathrm{test}}$. We create a Seurat object containing the counts $X^{\mathrm{train}}$, and 
preprocess  $X^{\mathrm{train}}$ and apply \texttt{FindClusters()} with the same arguments as above. We then add the $X^\mathrm{test}$ counts to the \texttt{Seurat} object as an additional assay, and we run \texttt{FindMarkers()} with the same arguments as above but with the additional argument \texttt{assay="test"}. We save the p-values for each gene returned by the \texttt{FindMarkers()} function. 
\item \textbf{Monocle3 (double dipping, as in the Monocle3 tutorial): } We first convert the data into a \texttt{cell\_data\_set} object. We then run the required preprocessing steps, and then run the functions \texttt{learn\_graph()} and \texttt{order\_cells()} to estimate pseudotime. We run all functions with their default settings, and for \texttt{order\_cells()} we set the root cell to be the first cell in the dataset. Finally, we run \texttt{graph\_test()} and save the Wilcoxon rank-sum p-values returned for each gene. 
\item \textbf{Monocle3 (count splitting): }	We first perform count splitting on the data to obtain $X^{\mathrm{train}}$ and $X^{\mathrm{test}}$. We create a \texttt{cell\_data\_set} object containing $X^{\mathrm{train}}$. We preprocess $X^{\mathrm{train}}$ and then run \texttt{learn\_graph()} and \texttt{order\_cells()} on the training object, with the same arguments as above. We then create a new \texttt{cell\_data\_set} object that is a copy of the training set object, and thus stores the same graph and pseudotime information. However, we update the \texttt{counts} attribute of this dataset to store $X^{\mathrm{test}}$. We then run \texttt{graph\_test()} on this new object and save the p-value returned for each gene. 
\end{itemize}

\begin{figure}
\includegraphics[width=\textwidth]{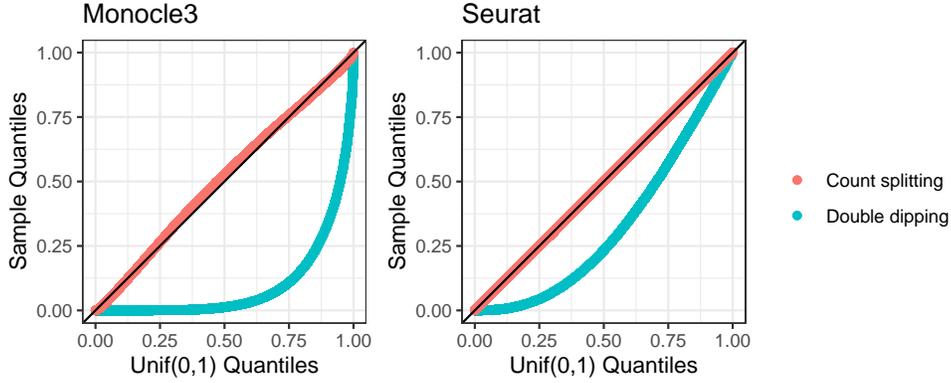}
\caption{Uniform QQ-plots of p-values under a global null, obtained using four different methods. }
\label{figure_monocle_seurat}
\end{figure}

Figure~\ref{figure_monocle_seurat} shows uniform QQ plots obtained from each of the four methods above. For both \texttt{Monocle3} and \texttt{Seurat}, the pipeline that uses the same data for clustering or pseudotime estimation and differential expression analysis leads to p-values that are too small, while the count splitting pipeline yields p-values that are uniformly distributed. This demonstrates that estimating a latent variable and testing for differential expression on the same data leads to invalid p-values. 

Code to reproduce Figure~\ref{figure_monocle_seurat} is available on our \texttt{R} package website at \texttt{anna-neufeld.github.io/countsplit}.

\section{Connections to \cite{batson2019molecular}}
\label{appendix_mcv}

To evaluate the goodness-of-fit of low-rank approximations of scRNA-seq expression matrices, with the ultimate goal of selecting optimal hyperparameters (such as the number of principal components to keep), \cite{batson2019molecular}
seek to obtain independent training and test sets. They propose \emph{molecular cross validation} (MCV), a method for obtaining such sets under the assumption that $X_{ij} \sim \mathrm{Binomial}(\Omega_{ij}, p_i)$, where $\Omega_{ij}$ is the true number of mRNA molecules for gene $j$ in cell $i$ and $p_i$ is the probability that a molecule in cell $i$ is observed. The splitting method involves three steps:
\begin{enumerate}
\item $X_{ij}^\mathrm{train} \sim \mathrm{Binomial}\left(X_{ij}, \epsilon \right)$
\item $X_{ij}^\mathrm{both} \sim \mathrm{Binomial}\left(X_{ij}^\mathrm{train}, p_i{''}\right)$
\item $X_{ij}^\mathrm{test} = X_{ij}-X_{ij}^\mathrm{train}+X_{ij}^\mathrm{both}$.
\end{enumerate}
Under the binomial assumption, if $p_i{''}$ is chosen appropriately, then $X_{ij}^\mathrm{train}$ and $X_{ij}^\mathrm{test}$ are independent. Unfortunately, choosing the appropriate $p_i''$ requires knowledge of the parameter $p_i$. 

As $p_i$ is typically unknown in practice, the authors suggest a simplification of the three-step process above in which $X_{ij}^\mathrm{both}$ is taken to be $0$. Under this simplification, MCV becomes identical to count splitting (Step 0 of Algorithm~\ref{alg_countsplit} in the main text). 

The authors' explanation for setting $X_{ij}^\mathrm{both}=0$ is as follows. They imagine that  $X_{ij}$ counts the total number of unique molecules seen through two separate, shallower sequencing experiments (which gave rise to $X_{ij}^\mathrm{train}$ and $X_{ij}^\mathrm{test}$). In this framework, $X_{ij}^\mathrm{both}$ denotes the molecules that were counted by both experiments and thus need to be subtracted out.  The authors note that, since scRNA-seq experiments tend to be quite shallow (i.e. the values of $p_i$ are small), the overlap between two even shallower experiments is likely to be negligible. We provide an alternate justification: when the $p_i$ are small and the $\Omega_{ij}$ are large, then the Poisson approximation to the binomial distribution holds, and so independence is given by Proposition~\ref{prop_independence}. 

Thus, while developed for a different goal and under different assumptions, the procedure of \cite{batson2019molecular} provides some justification for count splitting under a binomial assumption.

\section{Simulation comparing count splitting to selective inference}
\label{appendix_selectivecluster}

\cite{gao2020selective} provide a selective inference approach for testing the null hypothesis that two clusters 
estimated via hierarchical clustering have the 
same mean vector, under an assumption of multivariate normality. More specifically, they assume
\begin{equation}
\bold{X} \sim \mathcal{M}\mathcal{N}_{n \times q}\left( \bold{\mu}, \bold{I}_n, \sigma^2 \bold{I}_q \right),	
\end{equation}
and seek to test
\begin{equation}
\label{eq_selecnull}
H_0: \bar{\mu}_{\hat{\mathcal{C}}_1} = \bar{\mu}_{\hat{\mathcal{C}}_2},
\end{equation}
where $\hat{\mathcal{C}}_1$ and $\hat{\mathcal{C}}_2$ index the observations assigned to the two clusters and where, for any $G \subseteq \{1,\ldots,n\}$, we define
$
\bar{\mu}_G = \frac{1}{|i \in G|} \sum_{i \in G} \mu_i.
$
A naive Wald test for this null hypothesis does not control the Type 1 error rate because it ignores the fact that $\hat{\mathcal{C}}_1$ and $\hat{\mathcal{C}}_2$ are functions of the data $X$. \cite{gao2020selective} propose a selective Z-test that conditions on,
among other things, the event that $\hat{\mathcal{C}}_1$ and $\hat{\mathcal{C}}_2$ were output by the hierarchical clustering algorithm. As shown in their paper, this method controls the Type 1 error rate when the data truly come from a multivariate normal distribution. 

While not the focus of this paper, count splitting can also test the null hypothesis in \eqref{eq_selecnull} using a modification of Algorithm~\ref{alg_countsplit} from the main text. 

\begin{figure}
\centering
\includegraphics[width=0.6\textwidth]{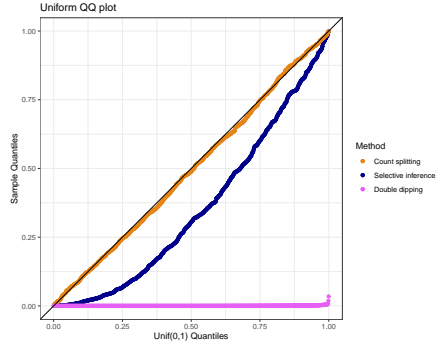}	
\caption{Uniform QQ-plot of p-values under a global null. Count splitting controls the Type 1 error while selective inference does not.} 
\label{fig_selective}
\end{figure}

To compare count splitting to selective inference in terms of Type 1 error rate control in this setting, we generate 1000 datasets where $n=200, p=10$ and $X_{ij} \sim \text{Poisson}\left(5\right)$. For each dataset, we do the following:
\begin{list}{}{}
\item \textbf{Double dipping:} Run hierarchical clustering with average linkage on $\log(X+1)$ to obtain two clusters $\hat{\mathcal{C}}_1$ and $\hat{\mathcal{C}}_2$. Use the naive test that double dips described in \cite{gao2020selective} on $\log(X+1)$ to test $H_0: \bar{\mu}_{\hat{\mathcal{C}}_1} = \bar{\mu}_{\hat{\mathcal{C}}_2}$. For this Wald test, estimate $\sigma$ using the ``conservative estimate" from \cite{gao2020selective}. 
\item \textbf{Selective:} Run hierarchical clustering with average linkage on $\log(X+1)$ to obtain two clusters $\hat{\mathcal{C}}_1$ and $\hat{\mathcal{C}}_2$. Use the selective test of \cite{gao2020selective} to test $H_0: \bar{\mu}_{\hat{\mathcal{C}}_1} = \bar{\mu}_{\hat{\mathcal{C}}_2}$. Estimate $\sigma$ using the ``conservative estimate" from \cite{gao2020selective}. 
\item \textbf{Count Split:} Run Step 0 (count splitting) of Algorithm~\ref{alg_countsplit} with $\epsilon=0.5$ to obtain $X^{\mathrm{train}}$ and $X^\mathrm{test}$. Run hierarchical clustering with average linkage on $\log(X^{\mathrm{train}}+1)$ to assign each cell to $\hat{\mathcal{C}}_1$ or $\hat{\mathcal{C}}_2$. Use the Wald test described in \cite{gao2020selective} on $\log(X^\mathrm{test}+1)$ to test $H_0: \bar{\mu}_{\hat{\mathcal{C}}_1} = \bar{\mu}_{\hat{\mathcal{C}}_2}$. For this Wald test, estimate $\sigma$ using the ``conservative estimate" from \cite{gao2020selective}, computed on the test set.
\end{list}
The results of this experiment are shown in Figure~\ref{fig_selective}. While selective inference improves upon the naive method that double dips, Figure~\ref{fig_selective} shows that it fails to control the Type 1 error rate for log-transformed Poisson data. This illustrates why the dependence of selective inference on a normality assumption makes it inadequate for scRNA-seq applications. 

\section{Implementation details for Figure~\ref{fig_null_qqs}(a)}
\label{appendix_fig1help}

In Figure~\ref{fig_null_qqs}(a), the double dipping method, cell splitting, and count splitting are all carried out as specified in Sections~\ref{section_existingmethods} and \ref{section_method}. The p-values reported are the default p-values returned by the \texttt{glm} function in \texttt{R} (Wald p-values). 

The \texttt{jackstraw} \texttt{R} package \citep{jackstrawRpackage} does not allow for fitting GLMs or for arbitrary latent variable estimation techniques. To obtain the jackstraw results for Figure~\ref{fig_null_qqs}(a), we implement the algorithm described in Section~\ref{subsec_jackstraw}, which is similar to the proposal of \cite{chung2015statistical} but allows for a GLM and an arbitrary latent variable estimation technique. We set $b=100$ and $s=10$ for our implementation. 

The \texttt{pseudotimeDE} \texttt{R} package \citep{pseudoRpackage} does allow for an arbitrary pseudotime estimation technique, but it fits negative binomial generalized additive models (GAMs) by default, and conducts likelihood ratio tests rather than Wald tests. To remain true to \cite{song2021pseudotimede}, the p-values for pseudotimeDE that are shown in Figure~\ref{fig_null_qqs}(a) are the default p-values returned by their \texttt{R} package (i.e. from GAMs and not GLMs). We estimate pseudotime using the first principal component of the log transformed matrix (as for the other methods). We then pass this estimate of pseudotime into the \texttt{R} package's \texttt{runPseudotimeDE()} function with its default settings. To compute each empirical p-value, we use $B=100$ subsets of the data, each containing 80\% of the observations.

\section{Proof of Propositions from Section~\ref{section_method}}

\subsection{Proof of Proposition~\ref{prop_negbincor}}
\label{appendix_negbincor}

Note that $\bold{X}_{ij}^\mathrm{train} \mid \{\bold{X}_{ij}=X_{ij}\} \sim \text{Binomial}(X_{ij}, \epsilon)$ and $\bold{X}_{ij}^\mathrm{test} \mid \{ \bold{X}_{ij} = X_{ij} \} \sim \text{Binomial}(X_{ij}, 1-\epsilon)$. The first statement is by construction and the second follows from swapping the roles of successes and failures in the binomial experiment. We now derive the marginal variance of each distribution from the known conditional mean and variance:
\begin{align*}
	\Var(\bold{X}^\mathrm{test}_{ij}) =& \E\left[ \text{Var}\left[\bold{X}^\mathrm{test}_{ij} \mid \bold{X}_{ij}
	 \right]\right] + \text{Var}\left[ \E\left[ \bold{X}^\mathrm{test}_{ij} \mid \bold{X}_{ij}
	  \right] \right] \\
	  &=\E\left[ (1-\epsilon)\epsilon \bold{X}_{ij} \right]  + \Var\left[ (1-\epsilon) \bold{X}_{ij} \right] \\
	  &= (1-\epsilon)\epsilon \Lambda_{ij}  + (1-\epsilon) ^2 \left( \frac{\Lambda_{ij}^2}{b_j} + \Lambda_{ij} \right) \\
	    &= (1-\epsilon) \Lambda_{ij} +  \frac{(1-\epsilon) ^2\Lambda_{ij}^2}{b_j},  \\
	 \Var(\bold{X}^\mathrm{train}_{ij}) =& \E\left[ \Var\left[\bold{X}^\mathrm{train}_{ij} \mid \bold{X}_{ij}
	 \right]\right] + \Var\left[ \E\left[ \bold{X}^\mathrm{train}_{ij} \mid \bold{X}_{ij}
	  \right] \right] \\
	  &= (1-\epsilon)\epsilon \Lambda_{ij}  + \epsilon ^2 \left( \frac{\Lambda_{ij}^2}{b_j} + \Lambda_{ij} \right)\\
	 &= \epsilon \Lambda_{ij}  + \frac{ \epsilon ^2 \Lambda_{ij}^2}{b_j}.
\end{align*}
Next note that 
$$
 \text{Var}(\bold{X}^\mathrm{train}_{ij}) + \text{Var}(\bold{X}^\mathrm{test}_{ij}) = \Lambda_{ij} +  \frac{(\epsilon^2 + (1-\epsilon)^2)\Lambda_{ij}^2}{b_j},
$$
and that since $\bold{X}_{ij} = \bold{X}^\mathrm{train}_{ij} + \bold{X}^\mathrm{test}_{ij}$, 
\begin{align*}
\mathrm{Cov}\left( \bold{X}_{ij}^\mathrm{train}, \bold{X}_{ij}^\mathrm{test}\right)	&= \frac{1}{2}\left( \Var\left(\bold{X}_{ij}\right) - \Var\left(\bold{X}_{ij}^\mathrm{train}\right) - \Var\left(\bold{X}_{ij}^\mathrm{test}\right) \right) \\
&= \frac{1}{2}\left(  \Lambda_{ij} + \frac{\Lambda_{ij}^2}{b_j} -  \Lambda_{ij} -  \frac{(\epsilon^2 + (1-\epsilon)^2)\Lambda_{ij}^2}{b_j}
\right) \\
&= \frac{1}{2}\left( \frac{2(\epsilon(1-\epsilon)) \Lambda_{ij}^2}{b_j}\right) = \frac{\epsilon(1-\epsilon) \Lambda_{ij}^2}{b_j}.
\end{align*}
Finally, to compute correlation, we divide by the covariance by
\small
\begin{align*}
\sqrt{\Var(\bold{X}_{ij}^\mathrm{train})\Var(\bold{X}_{ij}^\mathrm{test})} &= \sqrt{\left((1-\epsilon) \Lambda_{ij}  + \frac{ (1-\epsilon)^2 \Lambda_{ij}^2}{b_j} \right) 
\left(\epsilon \Lambda_{ij}  + \frac{ \epsilon ^2 \Lambda_{ij}^2}{b_j} \right)}\\
&= \sqrt{(1-\epsilon)\epsilon\Lambda_{ij}^2  + \frac{ \left[ \epsilon (1-\epsilon) \right] \Lambda_{ij}^3}{b_j} + \frac{ \epsilon^2 (1-\epsilon)^2 \Lambda_{ij}^4}{b_j^2}} \\
&= \frac{\Lambda_{ij}}{b_j}\epsilon(1-\epsilon) \sqrt{\frac{b_j^2}{\epsilon(1-\epsilon)} + \frac{b_j \Lambda_{ij}}{\epsilon(1-\epsilon)} + \Lambda_{ij}^2}. \\
\end{align*}
\normalsize
Putting it all together,
\begin{align*}
\mathrm{Cor}\left( \bold{X}_{ij}^\mathrm{train}, \bold{X}_{ij}^\mathrm{test}\right)	&= \frac{\frac{\epsilon(1-\epsilon) \Lambda_{ij}^2}{b_j}}{\frac{\Lambda_{ij}}{b_j}\epsilon(1-\epsilon) \sqrt{\frac{b_j^2}{\epsilon(1-\epsilon)} + \frac{b_j \Lambda_{ij}}{\epsilon(1-\epsilon)} + \Lambda_{ij}^2}} \\
&= \frac{\Lambda_{ij}}{\sqrt{\frac{b_j^2}{\epsilon(1-\epsilon)} + \frac{b_j \Lambda_{ij}}{\epsilon(1-\epsilon)} + \Lambda_{ij}^2}} \\
&= \frac{\sqrt{\epsilon(1-\epsilon)}}{\sqrt{\frac{b_j^2}{\Lambda_{ij}^2} + \frac{b_j }{\Lambda_{ij}} + \epsilon(1-\epsilon)}},
\end{align*}
as claimed in Proposition~\ref{prop_negbincor}.

\subsection{Proof of Proposition~\ref{prop_cor_poisson}}
\label{proof_cor_poisson}

Let $X_{ij} \sim \text{Poisson}\left( \gamma_i \Lambda_{ij} \right)$. First note that 
\begin{align*}
\text{Cov}(\bold{X}_{ij}, \bold{X}_{ij}^\mathrm{train}) &= \E\left[\bold{X}_{ij} E\left[X_{ij}^\mathrm{train} \mid X_{ij}\right]\right]	- \E[\bold{X}_{ij}]E[\bold{X}_{ij}^\mathrm{train}] \\
&= \epsilon \E[\bold{X}_{ij}^2] - \epsilon \gamma_i^2 \Lambda_{ij}^2  \\
&= \epsilon \left( \Var(\bold{X}_{ij}) + \E[\bold{X}_{ij}]^2 \right) - \epsilon \gamma_i^2 \Lambda_{ij}^2 \\
&=\epsilon \left( \gamma_i \Lambda_{ij} + \gamma_i^2 \Lambda_{ij}^2\right) - \epsilon \gamma_i^2 \Lambda_{ij}^2 \\
&= \epsilon \gamma_i \Lambda_{ij}.
\end{align*}
Next, note that $\mathrm{SD}(\bold{X}_{ij}) = \sqrt{ \gamma_i \Lambda_{ij}}$ and $\mathrm{SD}(\bold{X}_{ij}^\mathrm{train}) = \sqrt{ \epsilon\gamma_i \Lambda_{ij}}$. Thus, as claimed in Proposition~\ref{prop_cor_poisson},
\begin{align*}
\mathrm{Cor}(\bold{X}_{ij}, \bold{X}_{ij}^\mathrm{train}) &= \frac{\epsilon \gamma_i \Lambda_{ij}}{\gamma_i \Lambda_{ij} \sqrt{\epsilon}} = \sqrt{\epsilon}. 
\end{align*}

\subsection{Proof of Proposition~\ref{prop_power}}
\label{power_proof}

Let 
$
\bold{X}_{ij} \sim \mathrm{Poisson}\left(\gamma_i \exp(\beta_{0j} +\beta_{1j} L_i )\right).
$ 
The Fisher information matrix for the distribution of $\bold{X}_j$ with respect to $\beta_j = (\beta_{0j}, \beta_{1j})$ is given by
\begin{equation}
\label{fisher1}	
\mathcal{I}({\beta}_j) = [1_n~L]^T \text{diag}\left( E[\bold{X}_{j}]\right) [1_n~L].
\end{equation}
We define $\tilde{\beta}_0 = \log(1-\epsilon) + \beta_0$ and $\tilde{\beta}_1 = \beta_1$ such that, by Proposition~\ref{prop_independence}, we can write the distribution of $\bold{X}_{ij}^\mathrm{test}$ as $\mathrm{Poisson}\left(\gamma_i \exp(\tilde{\beta}_{0j} + \tilde{\beta}_{1j} L_i )\right)$. The Fisher Information matrix for the distribution of $\bold{X}_{j}^\mathrm{test}$ with respect to $\tilde{\beta}_j = (\tilde{\beta}_0, \tilde{\beta}_1)$ is given by
\begin{equation}
\label{fisher2}	
\mathcal{I}(\tilde{\beta}_j) = [1_n~L]^T \text{diag}\left( E[\bold{X}_{j}^\mathrm{test}\right) [1_n~L] =  (1-\epsilon)[1_n~L]^T \text{diag}\left( E[\bold{X}_{j}]\right) [1_n~L] =(1-\epsilon) \mathcal{I}({\beta}_j).
\end{equation}
If we regress $\bold{X}_j^\mathrm{test}$ on $L$, a 
Wald test for $H_0: \tilde{\beta}_1 = 0$ is based on the approximate large sample null distribution: 
$$
\hat{\beta}_1\left( L, \bold{X}_j^{\mathrm{test}}\right) \sim N\left(0, \left[I\left(\tilde{\beta}_j\right)^{-1}\right]_{22}\right).
$$
If we regress $\bold{X}_j$ on $L$, a Wald test for $H_0: {\beta}_1 = 0$ is based on the approximate large sample null distribution: 
$$
\hat \beta_1(L, {\bf X}_j) \sim N \left( 0, \left [\mathcal{I}\left({\beta}_j\right)^{-1} \right ]_{22} \right). 
$$
Thus, using \eqref{fisher2}, as claimed in Proposition~\ref{prop_power},
$$
\Var\left( \hat{\beta}_1\left( L, \bold{X}_j^{\mathrm{test}}\right)\right) \approx \frac{1}{1-\epsilon} \Var\left( \hat \beta_1(L, {\bf X}_j) \right).
$$
\section{Overdispersion in the cardiomyocyte data}
\label{appendix_realdata}

In Section~\ref{sec_realdata}, we fit Poisson GLMs to the cardiomyocyte differentiation data from \cite{elorbany2022single}. Here, we justify the use of Poisson GLMs (and the use of count splitting) by showing that the amount of estimated overdispersion is small. 

We carry out the following process using the $10,000$ cells from all 7 days of the differentiation protocol. We focus on the $p=2,500$ high variance genes that we analyze in Section~\ref{sec_realdata} of the main text (see Section~\ref{appendix_mymonocle}). 
\begin{enumerate}
\item Compute $\widehat{L}(X)$, using the function $\widehat{L}(\cdot)$ used in Section~\ref{sec_realdata} and elaborated on in Appendix~\ref{appendix_mymonocle}. 
\item For $j= 1,\ldots,p$, fit a negative binomial GLM using the \texttt{MASS} package in \texttt{R} to predict $X_j$ using $\widehat{L}(X)$. From each GLM, record the estimated overdispersion parameter $\hat{b}_j$, along with the predicted mean $\hat{\Lambda}_{ij}$ for $i=1,\ldots,n$ and $j=1,\ldots,p$. 
\item Make a histogram of the $n \times p$ values of $\frac{\hat{\Lambda}_{ij}}{\hat{b}_j}$. 
\end{enumerate}

We note that the process above double dips in the data, and so the estimated values $\hat{\Lambda}_{ij}$ and $\hat{b}_{j}$ may not be completely reliable. Here, we simply use them to understand the order of magnitude of the overdispersion.

\begin{figure}
\centering
\includegraphics[width=0.6\textwidth]{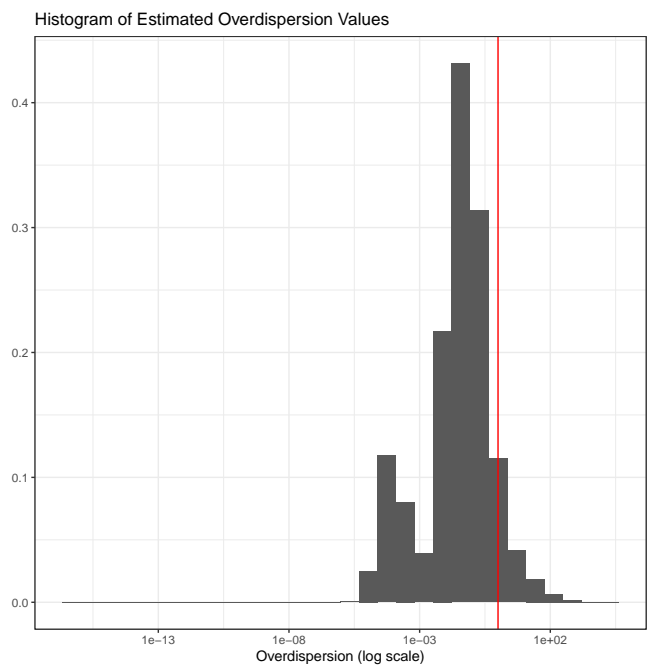}
\caption{Histogram of estimated values of $\frac{\Lambda_{ij}}{b_j}$ for the cardiomyocyte data. The vast majority of values are less than 1 (marked with a red vertical line).}	
\label{fig_real_overdisp}
\end{figure}

The resulting histogram is shown in Figure~\ref{fig_real_overdisp}. As noted in Section~\ref{subsec_overdisp}, the values of $\frac{\Lambda_{ij}}{b_j}$ indicate the amount of excess variance in the data. Figure~\ref{fig_real_overdisp} shows that, for the cardiomyocyte data, the vast majority of these values are less than 1. Thus, despite the presence of a few datapoints with very large values of overdispersion, we decided to fit Poisson GLMs (which are more numerically stable than negative binomial GLMs) for the analysis in Section~\ref{sec_realdata}. As shown in Figure~\ref{fig_null_qqs}(b) of the main text, count splitting performs reasonably well when $\frac{\Lambda_{ij}}{b_j} \leq 1$.  

\section{Details of pseudotime estimation from Section~\ref{sec_realdata}}
\label{appendix_mymonocle}

Let $M$ be the raw gene expression matrix. We now describe the procedure used to compute $\widehat{L}(\cdot)$ in Section~\ref{sec_realdata}. 

Steps 1--3 largely follow the pre-processing choices made in \cite{elorbany2022single}, whereas Steps 4--5 use the popular and user-friendly \texttt{Monocle3} \texttt{R} package for trajectory estimation and are motivated by the \texttt{Monocle3} vignettes \citep{monocle3vignette}. 

\begin{enumerate}
\item \textbf{Normalization: } We create the log-normalized matrix with the $(i,j)$th entry given by $\log\left( \frac{M_{ij}}{\hat{\gamma}_i(M)}+1 \right)$. All operations below are run on this log-normalized matrix. The $\hat{\gamma}(\cdot)$ function is the \texttt{librarySizeFactors()} function from the \texttt{scran} package \citep{scran}, which computes the row sums of $M$ and then scales them to have a geometric mean of 1.
\item \textbf{Estimate Cell Cycle:} Using the \texttt{tricycle} package \citep{tricycle}, estimate the cell cycle phase for each cell using the log-normalized version of $M$. 
\item \textbf{Feature selection: } We apply the \texttt{modelGeneVar()} function from the \texttt{scran} package. This function computes the mean and the variance of each log-normalized gene, and then fits a trend to these values. The residuals from this trend measure the biological component of variation for each gene. We then run the  \texttt{getTopHVGs()} from the \texttt{scran} package, which ranks the genes by this estimated biological component and computes a p-value to test the null hypothesis that this biological component is equal to 0. We select the genes whose biological component is significantly greater than 0 with a false discovery rate cutoff of 0.01. If more than 2500 genes are deemed significant, we retain the top 2500 of these selected genes for further analysis. While this feature selection is performed separately for the training dataset, test dataset, and full dataset inside of the $\widehat{L}(\cdot)$ function, the 2500 genes whose p-values are plotted in Figure~\ref{fig_realdata} of the main text are the top 2500 genes selected using the full dataset (rather than the training dataset or the test dataset). 
\item \textbf{Preprocessing and alignment and regressing out:} We reduce the dimension of the 
matrix output by the previous step (log-normalized with at most 2,500 features) by keeping only the top $100$ principal components of this matrix. Using the reduced dimension matrix, we then regress out technical sources of variation using the method of \cite{haghverdi2018batch}, as implemented in the \texttt{Monocle3} package in the function \texttt{align\_cds()}. We include cell line as the alignment group, and we additionally regress out the estimated cell cycle phase (from Step 2) and the proportion of mitochondrial reads. While this regression step is suggested in the \texttt{Monocle3} vignettes, regressing out the estimated cell cycle phase is perhaps less standard. Regressing out cell cycle is particularly important for the ``day 0 only" example, as we want to ensure that no true trajectory (including a trajectory through the cell cycle) is present in the ``day 0 only" data. 
\item \textbf{Monocle3 dimension reduction, graph embedding, and pseudotime calculation:} Next, we compute pseudotime using a sequence of functions from the \texttt{Monocle3} package. All functions are run with their default settings unless otherwise noted. The input to this step is the matrix output by the previous step. First, 
we run \texttt{reduce\_dimension()}, which takes the already dimension-reduced data and turns it into a two-dimensional UMAP representation. Next, we run \texttt{cluster\_cells()}. (While we do not wish to estimate clusters, this is a required step of the \texttt{Monocle3} pipeline.) We next run \texttt{learn\_graph()}, but we let \texttt{use\_partition=FALSE} so that \texttt{Monocle3} ignores cluster information and learns a single graph through all of the cells. The clustering and graph steps both happen in UMAP space. Finally, we run \texttt{order\_cells()}, which projects all cells onto the principal graph to obtain a single continuous trajectory. As \texttt{order\_cells()} requires a user-specified ``root" cell (the cell which will have a pseudotime of $0$), we choose the induced pluripotent stem cell (IPSC) that is closest to a vertex of the graph as the ``root" cell. The cell type labels (i.e. the labels that allow us to identify IPSC cells) were computed by \cite{elorbany2022single}. 
\end{enumerate}

\end{document}